%% file: main.tex
\documentclass[
  journal=pasa,
  manuscript=research-paper, 
  year=\the\year,
  volume=x,
]{cup-journal}

\usepackage{microtype,siunitx,booktabs}
\usepackage{amssymb}
\usepackage{gensymb}

\newcommand\hi{\mbox{\sc Hi}}
\newcommand\hii{\mbox{\sc Hii}}

\graphicspath{{./}{figures/}}

\title[SMC Absorption with GASKAP]{GASKAP-HI Pilot Survey Science III: An unbiased view of cold gas in the Small Magellanic Cloud}

\author{James~Dempsey}
\email{james.dempsey@anu.edu.au}
\affiliation{Research School of Astronomy and Astrophysics, 
The Australian National University, 
Canberra, ACT 2611, Australia}
\alsoaffiliation{CSIRO Information Management and Technology, 
GPO Box 1700 
Canberra, ACT 2601, Australia}

\author{N.~M.~McClure-Griffiths}
\affiliation{Research School of Astronomy and Astrophysics, 
The Australian National University,  
Canberra, ACT 2611, Australia}

\author{Claire~Murray}
\affiliation{Department of Physics \& Astronomy,
Johns Hopkins University,
3400 N. Charles Street, 
Baltimore, MD 21218}
\alsoaffiliation{Space Telescope Science Institute, 
3700 San Martin Drive, 
Baltimore, MD, 21218}

\author{John~M.~Dickey}
\affiliation{School of Natural Sciences, 
Private Bag 37, University of Tasmania, 
Hobart, TAS, 7001, Australia}

\author{Nickolas~M.~Pingel}
\affiliation{Research School of Astronomy and Astrophysics,
The Australian National University,
Canberra, ACT 2611, Australia}

\author{Katherine Jameson}
\affiliation{ATNF, CSIRO, Space and Astronomy, 26 Dick Perry Avenue, Kensington WA 6151, Australia}

\author{Helga D\'{e}nes}
\affiliation{ASTRON - The Netherlands Institute for Radio Astronomy,
7991 PD Dwingeloo, The Netherlands}

\author{Jacco~Th.~van~Loon}
\affiliation{Lennard-Jones Laboratories,
Keele University
ST5 5BG, UK}

\author{D.~Leahy}
\affiliation{Department of Physics and Astronomy, University of Calgary, Calgary, AB T2N 1N4, Canada}

\author{Min-Young~Lee}
\affiliation{Korea Astronomy and Space Science Institute,
776, Daedeokdae-ro, Yuseong-gu 
Daejeon 34055, Republic of Korea}

\author{S.~Stanimirovi\'{c}}
\affiliation{Department of Astronomy, University of Wisconsin–Madison 
475 North Charter Street, Madison, WI 53706-15821, USA}

\author{Shari~Breen}
\affiliation{SKA Observatory, Jodrell Bank, Lower Withington, Macclesfield, Cheshire SK11 9FT, UK}

\author{Frances Buckland-Willis}
\affiliation{AIM, CEA, CNRS, Universit\'e Paris-Saclay, Universit\'e Paris Diderot, Sorbonne Paris Cit\'e, F-91191 Gif-sur-Yvette, France}

\author{Steven J. Gibson}
\affiliation{Department of Physics and Astronomy,
Western Kentucky University,
Bowling Green, KY 42101, USA}

\author{Hiroshi~Imai}
\affiliation{Center for General Education, Comprehensive Institute of Education, Kagoshima University,
1-21-30 Korimoto, Kagoshima 890-0065, Japan}
\alsoaffiliation{Amanogawa Galaxy Astronomy Research Center, Graduate School of Science and Engineering, Kagoshima University,
1-21-30 Korimoto, Kagoshima 890-0065, Japan}

\author{Callum~Lynn}
\affiliation{Research School of Astronomy and Astrophysics, 
The Australian National University,  
Canberra, ACT 2611, Australia}

\author{C.~D.~Tremblay}
\affiliation{CSIRO Space and Astronomy, 
PO Box 1130, Bentley, WA 6102, Australia}

\doi{10.1017/pasa.\the\year.xxx}

\received {3 Dec 2021}
\revised  {10 Apr 2022}
\accepted {13 Apr 2021}
\published{dd Mmm YYYY}

\keywords{
Small Magellanic Cloud -- Interstellar line absorption -- Dwarf irregular Galaxies -- Interstellar medium -- Neutral hydrogen clouds
}

\begin{document}

\begin{abstract}

We present the first unbiased survey of neutral hydrogen (HI) absorption in the Small Magellanic Cloud (SMC).
The survey utilises pilot HI observations with the Australian Square Kilometre Array Pathfinder (ASKAP) telescope as part of the Galactic ASKAP HI (GASKAP-HI) project whose dataset has been processed with the GASKAP-HI absorption pipeline, also described here.
This dataset provides absorption spectra towards 229 continuum sources, a 275\% increase in the number of continuum sources previously published in the SMC region, as well as an improvement in the quality of absorption spectra over previous surveys of the SMC.
Our unbiased view, combined with the closely matched beam size between emission and absorption, reveals a lower cold gas faction (11\%) than the 2019 ATCA survey of the SMC and is more representative of the SMC as a whole.
We also find that the optical depth varies greatly between the SMC's bar and wing regions.
In the bar we find that the optical depth is generally low (correction factor to the optically thin column density assumption of $\mathcal{R}_{\rm HI} \sim 1.04$) but increases linearly with column density. 
In the wing however, there is a wide scatter in optical depth despite a tighter range of column densities. 
\end{abstract}

\section{Introduction}
\label{sec:intro}

Neutral hydrogen (\hi) exists in multiple phases in a galaxy's interstellar medium (ISM).
\hi\ is observed in two long-lived phases, the warm neutral medium (WNM, 5000-10,000 K) and cold neutral medium (CNM, 20-200K) \citep{McKee+77}.
In a pressure equilibrium these two phases will coexist, but that equilibrium is dependent on metallicity \citep{1995ApJ...443..152W,2003ApJ...587..278W,Bialy+2019}.
Turbulence and colliding flows can drive the formation of CNM \citep{Hennebelle+2007,Kim+2017}.
The CNM is a precursor to the formation of the dense cores of molecular hydrogen (H$_2$) from which stars form.
The fraction of \hi\ in the CNM state is an important metric for the efficiency of star formation and galaxy evolution, along with the molecular gas fraction \citep{Krumholz+2009,Kennicutt+2012}.

We can use the spin temperature ($T_{\rm S}$) of \hi\ to assess the fraction of cold gas.
The spin temperature of \hi\ is the excitation temperature of the \hi\ 21-cm spin-flip transition.
As the gas is thermalised by collisions in the dense CNM, the spin temperature will be equal to the kinetic temperature of the gas in this environment \citep{Field58}.
In the WNM, it will be a lower limit for the kinetic temperature \citep{Liszt01}.

By combining absorption observations with adjacent emission observations we are able to measure the spin temperature of the gas \citep{Murray+2015,Jameson+2019}.
\hi\ absorption against continuum sources allows us to directly detect the presence of cold gas clouds.
These clouds are otherwise difficult to detect as, due to their low spin temperature, they produce only weak 21-cm emission.

The Small Magellanic Cloud (SMC) is the perfect laboratory to study cold gas formation at high resolution in a low metallicity environment.
The SMC is a nearby ($61$ kpc; \citealt{Graczyk+2014}) low mass galaxy, part of the interacting Magellanic System, along with the Large Magellanic Cloud, the Leading Arm and the trailing Magellanic Stream.
The SMC has metallicity of ~0.2 solar \citep{Russell+1992} and thus will have a typical temperature range for the CNM of $50 \le {\rm T_S} \le 100$ K \citep{Bialy+2019}.

\cite{Dickey+00} presented the first survey of \hi\ absorption across the SMC, examining emission and absorption towards 32 continuum sources. 
They achieved optical depth noise levels of $ 0.05 \leq \sigma_{\tau} \leq 0.203$ at 0.825 km s$^{-1}$ spectral resolution and detected significant absorption in 13 of these spectra.
The sources were selected for their strong continuum flux.
\cite{Jameson+2019} built on this by examining emission and absorption towards 55 continuum sources in 21 fields.
These fields were selected for the greatest likelihood of finding absorption, with strong continuum flux ($S_{\rm cont} > 50$ mJy) and towards high column density regions ($N_{\rm HI} > 4 \times 10^{20}$ cm$^{-2}$).
In their $\ge 10$ hr dwell time per target they reached noise levels of $ 0.01 \leq \sigma_{\tau} \leq 1.28$ and detected absorption in 37 of the spectra.
Their spectra were imaged at 0.2 km s$^{-1}$ but smoothed to 0.6 km s$^{-1}$ for analysis.
Both surveys used the Australia Telescope Compact Array with baselines up to 6 km  (FWHM$_{\rm beam} \approx 5$ arcsec) for absorption and compared them to emission from the  \cite{Stanimirovic+99} survey of the SMC (FWHM$_{\rm beam} \approx 98$ arcsec).
Table \ref{tab:surveys} summarises the observational parameters of these surveys and the survey presented in this work.

\begin{table}
    \centering
    \begin{tabular}{lrrrrr}
    \hline \hline
    Survey & $\Delta v$ & FWHM & $\sigma_{\tau}$ & Num & Det.\footnote{\cite{Dickey+00} and \cite{Jameson+2019} were targeted surveys which affects the detection rate. }  \\
     &  (km s$^{-1}$)& (")  &  & Src & Rate   \\
    \hline
    \cite{Dickey+00} & 0.825 & 5 & 0.05--0.203 & 32 & 0.41  \\
    \cite{Jameson+2019} & 0.6 & 5 & 0.01--1.28 & 55 & 0.67  \\
    This work & 1 & 16 & 0.01--0.29 & 229 & 0.28  \\
    \hline
    \end{tabular}
    \caption{Comparison of SMC absorption survey parameters}
    \label{tab:surveys}
\end{table}

Using the Australian Square Kilometre Array Pathfinder (ASKAP) telescope, the Galactic ASKAP (GASKAP; \citealt{2013PASA...30....3D}) survey will observe \hi\ and OH in the Galactic Plane and the Magellanic System with unprecedented detail.
The observations use high angular resolution (FWHM$_{\rm beam} \approx 16$ arcsec) and high spectral resolution (0.24 km s$^{-1}$ per channel).
Planned observations include long dwell times on the Magellanic Clouds and the low latitude Galactic Plane and shorter dwell times on the Magellanic Stream and Bridge.
The repeated observations of key fields will provide \hi\ absorption spectra with a flux density sensitivity of $\sigma_{\rm S} = 0.5 $ mJy \citep{2013PASA...30....3D}.
The survey is planned to cover 13,020 deg$^2$ in total.
With a rate of $\approx 10$ sources per square degree, up to 130,200 \hi\ absorption spectra are expected in the GASKAP-HI survey.
With such high volumes of spectra, a repeatable process is essential.

In this work we present the GASKAP \hi\ absorption pipeline and use it with the pilot phase I SMC \hi\ observations \citep[see][]{pingel2021} to explore the distribution of cold gas in the SMC in an unbiased way.
In section 2, we describe the GASKAP pilot observations of the SMC. 
In section 3, we present our \hi\ absorption pipeline along with the processing parameters used. 
We describe the observed absorption in the SMC and surrounds in section 4.
In section 5, we discuss the results and their implications.
Finally, in section 6, we summarise our findings.

\section{Observations}
\label{sec:obs}

The SMC was the first of three adjacent Magellanic fields targeted during the GASKAP Pilot Phase I observations. 
Two 12-hour observations of the field (ASKAP scheduling blocks 10941 and 10944) were taken in December 2019 using the standard GASKAP-HI observing configuration. 
The closepack-36 phased-array feed (PAF) footprint was used with a pitch of $0.9$ deg and 3 interleaves for even coverage across the 25 square degree field.
The field was centred on J2000 RA = 00$^{\rm h}$58$^{\rm m}$43.280$^{\rm s}$, Dec = $-$72$^{\rm d}$31$^{\rm m}$49.03$^{\rm s}$.
The zoom-16 mode, with $\Delta\nu = 1.15$ kHz, was used to provide a spectral resolution of $\Delta v \sim 0.24$ km s$^{-1}$.
The observed band covered 18.5 MHz centred on 1419.85 MHz with 15558 channels, however only the 2048 channels covering the Milky Way and SMC velocity ranges were processed.
The flagging was even across the field, providing a change in RMS across the field of $\leq 7.5 \%$. 

As noted in \cite{2013PASA...30....3D}, ASKAP's bimodal baseline distribution provides excellent capabilities for measuring both \hi\ emission and absorption. 
In these observations, there was good coverage of baselines longer than 2000m, which provide fine spatial resolution.
Baselines up to 6000m were present. 
The expected sensitivity from the combined observation is 3.3 mJy/beam after accounting for flagging and excluded baselines (see Section \ref{sec:imgParams}).

The data were flagged and calibrated using the standard ASKAPSoft pipeline  \citep{2021PASA...38....9H} with configuration suitable for wide field emission. 
A continuum image and a continuum source catalogue were also produced by the ASKAPSoft pipeline for each observation.
The observations and the initial processing with ASKAPSoft are described in further detail in \cite{pingel2021}.

\section{\hi\ Absorption Pipeline}
\label{sec:data}

To measure the absorption against continuum sources we need a spectral line cube for the region surrounding each source. 
We have two potential strategies to produce these source cubes:
\begin{enumerate}[label=\alph*)]
    \item producing a large, continuum included cube covering the entire field and then extracting sub-cubes for each source,  or
    \item producing sub-cubes for each source position directly from the measurement sets.
\end{enumerate}
Note, we cannot use the GASKAP emission cube as the continuum subtraction will limit our ability to accurately measure deep absorption, and the cube contains large-scale emission which will add noise to the absorption spectrum.
We have chosen to take approach (b) as we know in advance the positions we wish to image.
This approach also has several advantages for our use case. 
This saves us from imaging the unused regions ($> 99$\%) of the cube not sampled by the sparse and small continuum sources.
Moreover, we are able to set the phase-centre to the source position for each sub-cube, thus avoiding $w$-term effects which can impact wide fields \citep{1992a&a...261..353c}.
Finally we are able to optimise the cube production to obtain the most accurate absorption spectra possible.

We have developed a pipeline \citep{james_dempsey_2022_6388108} to produce \hi\ absorption spectra for a large number of sources from the calibrated ASKAP measurement sets.
As GASKAP is a wide field survey, and not targeted at specific sources, we have the opportunity to take an unbiased sample of the cold \hi\ absorption in every field. 
We obtain absorption spectra towards all continuum sources above a given brightness (e.g. $S_{\rm cont} \geq 15$ mJy).
For each source, we produce a small cube around the source, sufficient to include all continuum emission from the source. 
From this we extract an integrated absorption spectrum. 
We also measure the emission surrounding the source from the GASKAP emission cube for the field \citep{pingel2021}.
Finally, we then process all spectra to find any significant absorption detections and produce catalogues of spectra and absorption features.

The pipeline takes as input:
\begin{enumerate}
    \item A continuum source catalogue, the ASKAP Selavy \citep{2012PASA...29..371W} component catalogue
    \item A collection of calibrated ASKAP beam measurement sets. 
\end{enumerate}
The pipeline is split into three phases: preparation, imaging, and spectra extraction, each described below.

\subsection{Preparation Phase}
\label{subsec:prepPhase}

In the preparation phase we identify the target sources and the input data for the pipeline run. 
We take the continuum component catalogue produced as part of the ASKAPSoft processing of the field \citep{2021PASA...38....9H}.
We designate all sources with $S_{\rm cont} \geq 15$ mJy as target sources.
A typical GASKAP-HI observation for 20 hrs has a $\sigma_{\rm S} = 1.89$ mJy/beam \citep{pingel2021}, and after the baseline cutoff described in Section~\ref{sec:imaging} $\sigma_{\rm S} = 2.80$ mJy/beam.
For sources with $S_{\rm cont} = 15$ mJy we have a theoretical optical depth noise level of $\sigma_{\rm cont} = 0.19$.
Examining the 47 sources near this threshold ($15 \leq {\rm S_{cont}} \le 17$ mJy) in the SMC pilot data, we see a median optical depth noise level of $\sigma_{\rm cont,median} = 0.23$ with a range of $0.07 \leq \sigma_{\rm cont} \leq 1.34$, showing that we are at the limit of useful spectra in these data.
For the 200 hr integrations on the Magellanic Clouds planned in the full survey, we could expect to push this limit down to as low as $S_{\rm cont} \sim 5$ mJy with similar optical depth noise levels.

For each target source we use all beams within 0.55 deg of the source (or 0.8 deg for edge sources where no beams are closer).
From these we produce a list of all target sources and the beams which will be used to produce the cutout for each source.

\subsection{Imaging Phase}
\label{sec:imaging}

The imaging phase is the most computationally intensive. 
As discussed earlier, we produce a sub-cube around each source. We process this phase in parallel with a job per sub-cube.
In the pipeline, we 
dynamically schedule the sub-cube extraction jobs to minimise competing use of measurement sets and thus improve overall I/O throughput while providing high parallelism within the resources of the compute infrastructure.

For each cube we use the CASA \citep{2007aspc..376..127m} task {\tt tclean} to image a 50 arcsec x 50 arcsec region around the source at 1 arcsec per pixel and with 0.24 km~s$^{-1}$ LSRK velocity resolution (the native spectral resolution).
This images the beam measurement sets selected for each source in the preparation phase.
We use a  1.5 k$\lambda$ baseline length cutoff, natural weighting, primary beam correction and light cleaning (1000 total iterations), as discussed in Section~\ref{sec:imgParams}.
This gives a typical synthesised beam size of $16$ x $14$ arcsec$^2$.
We use a primary beam model of a 12 m dish with a 0.75 m blockage.

\subsubsection{Imaging Parameters for GASKAP Absorption}
\label{sec:imgParams}

\begin{figure}[ht]
    \centering
    \includegraphics[scale=0.8]{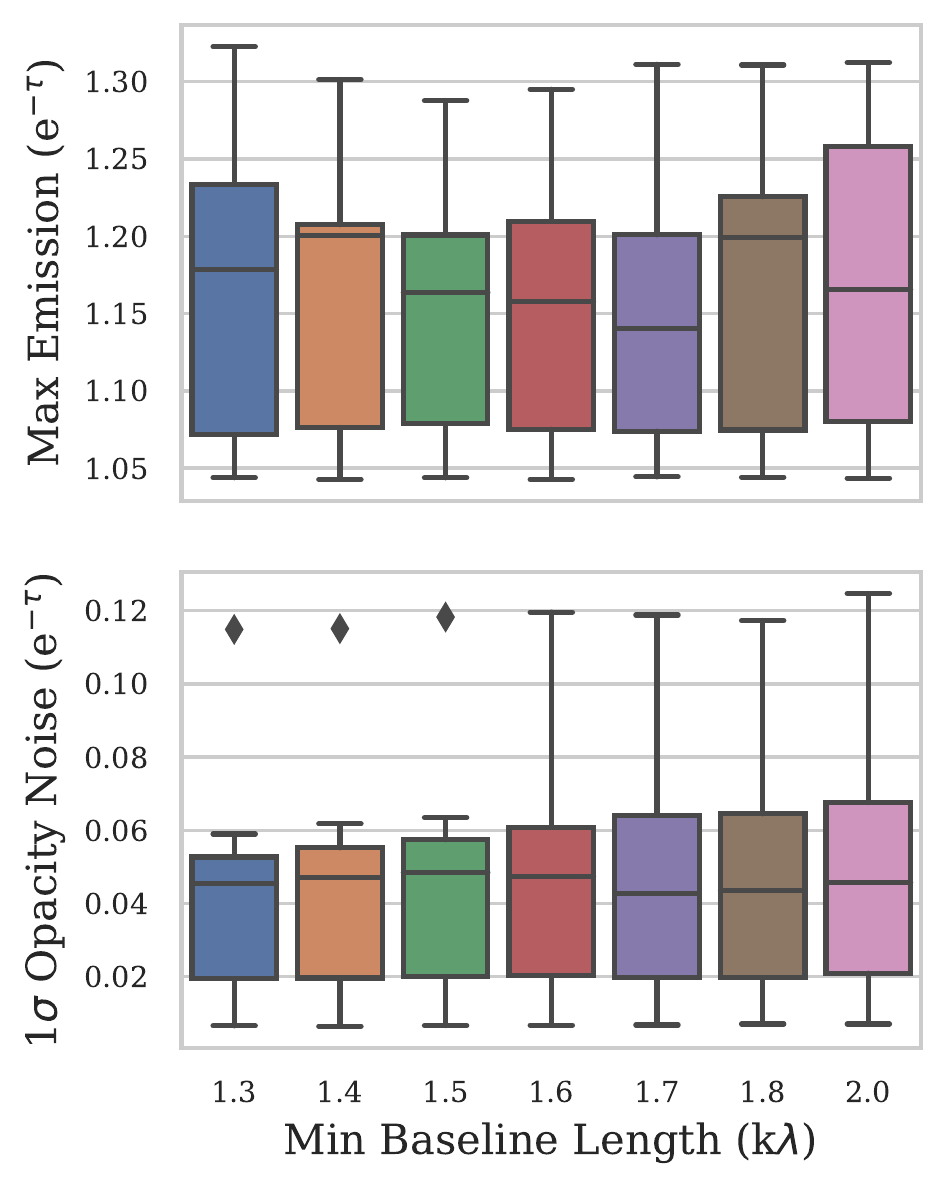}
    \caption{Box-plot comparison of maximum levels of noise from emission (where $e^{-\tau} > 1$; top) and optical depth noise (bottom) for each of the sample spectra for a range of baseline length cutoffs. In these plots the central horizontal line is the median, the ends of the box are the 25th and 75th percentiles and the top and bottom lines show the maximum and minimum values respectively. Diamonds show outliers based on their distance from the interquartile range.}
    \label{fig:baselineTest}
\end{figure}

\begin{figure}[ht]
    \centering
    \includegraphics[scale=0.8]{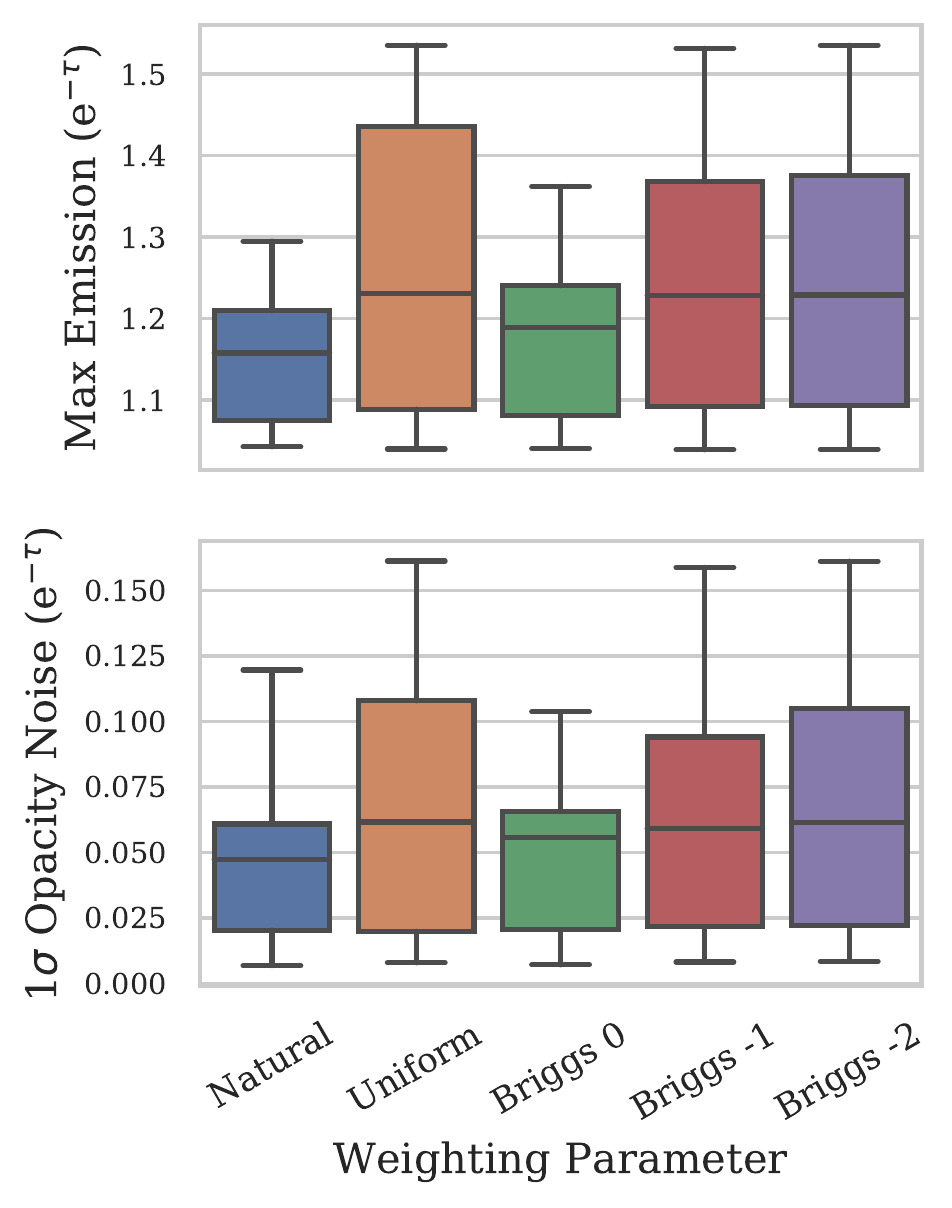}
    \caption{Box-plot comparison of maximum levels of noise from emission (where $e^{-\tau} > 1$; top) and optical depth noise (bottom) for each of the sample spectra for a range of weighting parameters. See Fig.~\ref{fig:baselineTest} for details of the ranges.}
    \label{fig:weightingTest}
\end{figure}

For absorption studies we wish to exclude emission from the spectra while maintaining high signal-to-noise ratios. 
In this pipeline we have achieved that primarily by setting a minimum baseline length when producing the cutout cube. 
Figure \ref{fig:baselineTest} shows the results of our trials of a range of baseline length cutoffs on a sample of seven spectra selected to represent the diverse range of spectra seen in the full dataset.
Based on these trials, we found that the optimum balance of reduced emission and reduced noise came from a 1.5 k$\lambda$ (315 m) baseline length cutoff.
This retains 574 of the 630 ASKAP baselines but provides sensitivity only to features of 
$\lesssim3$ arcmin in size.
This is well suited for analysis of compact extra-Galactic sources, but means these data products would not be suitable for analysis of larger Galactic structures such as supernova remnants and \hii\ regions. 

Similarly, we tested a variety of weighting parameters, as shown in Figure \ref{fig:weightingTest}. 
Based on these results we
found `natural' weighting produced cutout cubes with the lowest optical depth noise and emission noise.
Additionally, we used primary beam correction and light cleaning (1000 total iterations) when imaging the data.

\subsection{Spectra Extraction Phase}
\label{sec:specExtract}

From the cutout cubes for each source, we extract absorption spectra for every target source.
We use the technique developed in \citet{2020MNRAS.496..913D}, but we describe the process here for completeness.
The continuum source catalogue defines a source ellipse for each component.
We combine all pixels within this source ellipse to produce the spectrum.
We define a line-free region of the cube (with a velocity range $-100 \leq $v$_{\rm LSRK} \leq -60$ km~s$^{-1}$ for the SMC) and measure the mean brightness within this range for each pixel.
We then weight each pixel's contribution to the spectrum by the square of the mean brightness, as described in \citet{1992ApJ...385..501D}.
Lastly, we divide the combined spectrum by its mean brightness within the line-free region to produce the absorption spectrum, $e^{-\tau}$.

We estimate the noise in the spectrum using a combination of the noise in the off-line region and emission in the primary beam of the dish, as described in  \cite{Jameson+2019}.
The standard deviation of the spectrum in the line-free region is taken as the base noise for the spectrum. 
In order to model the increase in system temperature due to emission received by the antenna at different frequencies, we then measure the emission level in the ASKAP primary beam of ~62 arcmin using data from the Parkes Galactic All-Sky Survey (GASS; \citealt{2009ApJS..181..398M,2015A&A...578A..78K}). 
We average the GASS emission across a 7 pixel (33 arcmin) radius annulus centred on the source position, with a 1 pixel exclusion at the centre.
The 1$\sigma$ noise envelope for the GASKAP absorption spectrum is then calculated as:
\begin{equation}
\sigma_{\tau}(v) = \sigma_{\rm cont} \frac{T_{\rm sys} + \eta_{\rm ant}T_{\rm em}(v)}{T_{sys}},
\end{equation}
where the system temperature $T_{\rm sys} = 50$ K and antenna efficiency $\eta_{\rm ant} = 0.67$ \citep{2021PASA...38....9H}, $\sigma_{\rm cont}$ is the standard deviation of the line-free region of the spectrum, and the mean emission $T_{\rm em}(v)$ at each velocity step is as measured in GASS.

We identify any absorption features in the spectra during this phase.
We classify absorption features as those having one channel of $3 \sigma$ absorption and an adjacent channel of $\geq 2.8 \sigma$. 
We expand the feature to include any adjacent channels of at least $2.8 \sigma$ significance. 
These criteria were chosen empirically to detect even shallower absorption whilst still avoiding noise spikes being detected as features.
Statistically, these criteria give an average of less than one false positive absorption feature per $\approx 370$ spectra.

With a large number of spectra expected in each field, it is important to consistently identify which spectra are of sufficient quality to be analysed. 
We use a subset of the \citet{2014ApJS..211...29B} rating system to classify the spectra quality level from A to D.
Rating A spectra pass all tests, while each spectrum's rating is reduced by one step for each failed test until rating D spectra fail all tests.
The tests described in Brown et al. aim to flag unphysical or noisy spectra.

The three tests we use are:
\begin{itemize}
    \item Continuum noise - The $1\sigma$ noise level ($\sigma_{\rm cont}$) in optical depth ($e^{-\tau}$) must be less than $1/3$
    \item Max signal to max noise - The ratio of the deepest absorption to the highest emission noise (i.e. $e^{-\tau} > 1$) in the spectrum: $(1-{\rm min}( e^{-\tau}))/({\rm max}( e^{-\tau}) -1) \ge 3$
    \item Optical depth range - The range from the deepest absorption to highest emission noise in the spectrum: $ {\rm max}(e^{-\tau}) - {\rm min}(e^{-\tau}) < 1.5$.
\end{itemize}
We do not use the other two Brown et al. tests as they are not applicable to our spectral extraction method.

To make measurements of spin temperature it is also necessary to have an emission spectrum for each source.
The emission spectra are calculated from the full field continuum-subtracted GASKAP emission cube produced for the same observations \citep{pingel2021}.
We calculate the emission spectrum of a source as the mean value of an annulus of radius 56 arcsec (8 pixels, $\approx 2$ beam widths) around each source with a central exclusion of 28 arcsec (4 pixels, $\approx 1$ beam width).
This allows us to measure the mean emission in the close vicinity of the source without measuring the source itself, in all but the most extended cases.
The high resolution gives us as close an approximation for the emission in the line of sight of the source as possible.
We use the standard deviation of the annulus as the emission noise level.

The pipeline produces a series of outputs describing the spectra towards each source.
For each source we produce a spectrum votable file with the absorption and, where available, the emission spectra, along with the noise in both the absorption and emission spectra.
We output catalogues of the spectra, (see Table \ref{tab:spectra}), as well as the absorption features detected (see Table \ref{tab:absorption}).

\section{SMC \hi\ Absorption}
\label{sec:absorption}

\begin{figure*}[ht]
    \centering
    \includegraphics[width=\linewidth]{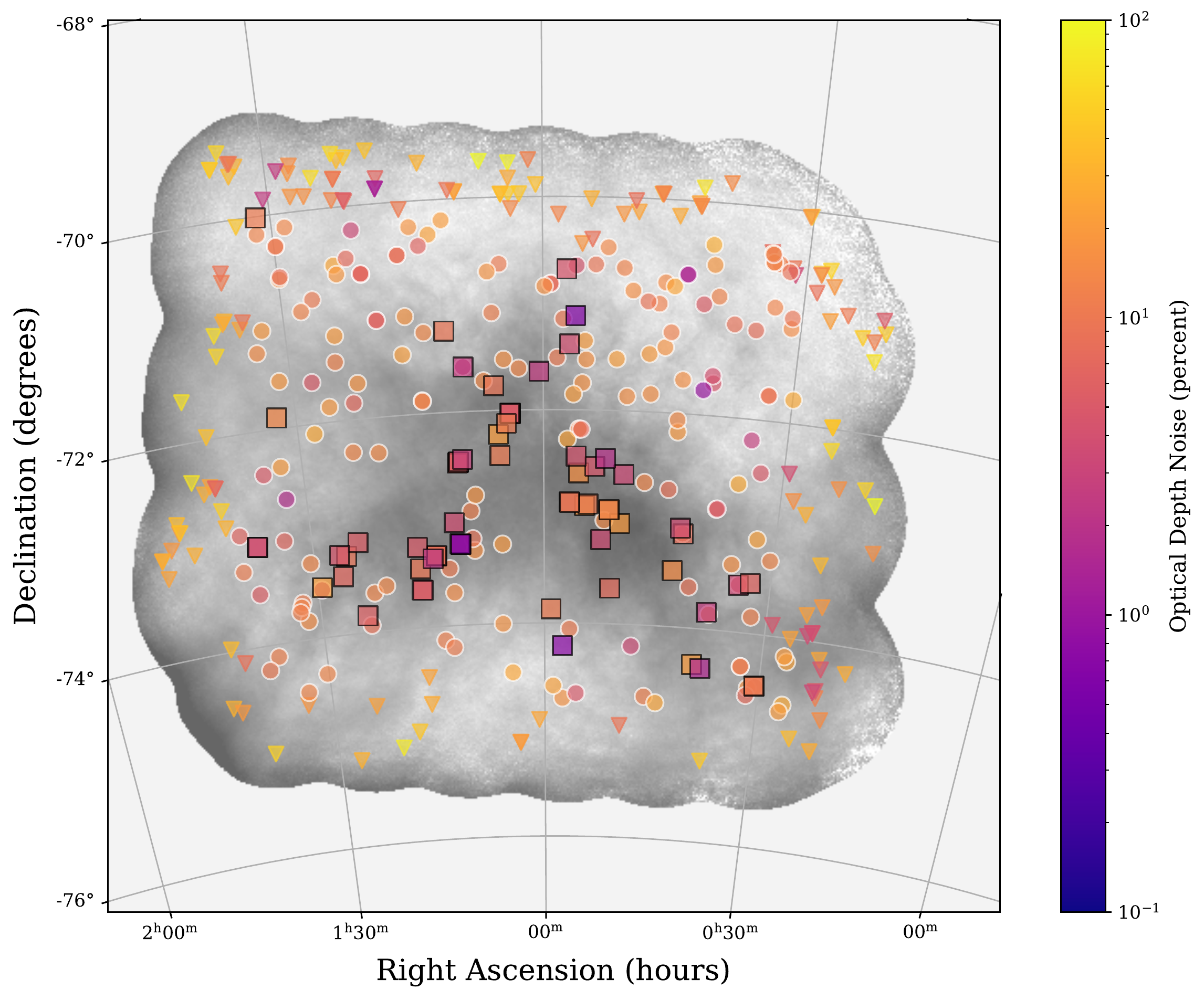}
    \caption{Distribution of continuum sources showing their optical depth noise against the SMC \hi\ column density map from GASKAP \citep{pingel2021}.
    Triangles are sources excluded due to either high noise or being on the edges of the cube, squares are sources against which absorption was detected, and circles are other sources.
    Darker colours indicate lower optical depth noise.}
    \label{fig:allsources}  
\end{figure*}

For each of the 373 continuum sources which satisfy our criteria of peak flux density $\ge 15$ mJy (see Sec.~\ref{subsec:prepPhase}), we produced an absorption spectrum using the process described in Section \ref{sec:specExtract}.
However, we averaged the data to 1 km s$^{-1}$ spectral resolution in the imaging phase to reduce noise.
We selected the sources from the Selavy \citep{2012PASA...29..371W} continuum source catalogue from the SB 10941 observation. 
We then excluded any sources with $\sigma_{\rm cont} > 0.3$ or at the edges of the field where beam power $< 80\%$, leaving 229 sources. 
The source density is 10 sources per square degree, a rate which we expect to be replicated in most GASKAP fields of similar noise.
The distribution of all sources by noise is shown in Figure~\ref{fig:allsources}.
Notably, all of the excluded noisy sources are towards the edges of the field.

\begin{figure}[ht]
    \centering
    \includegraphics[width=\linewidth]{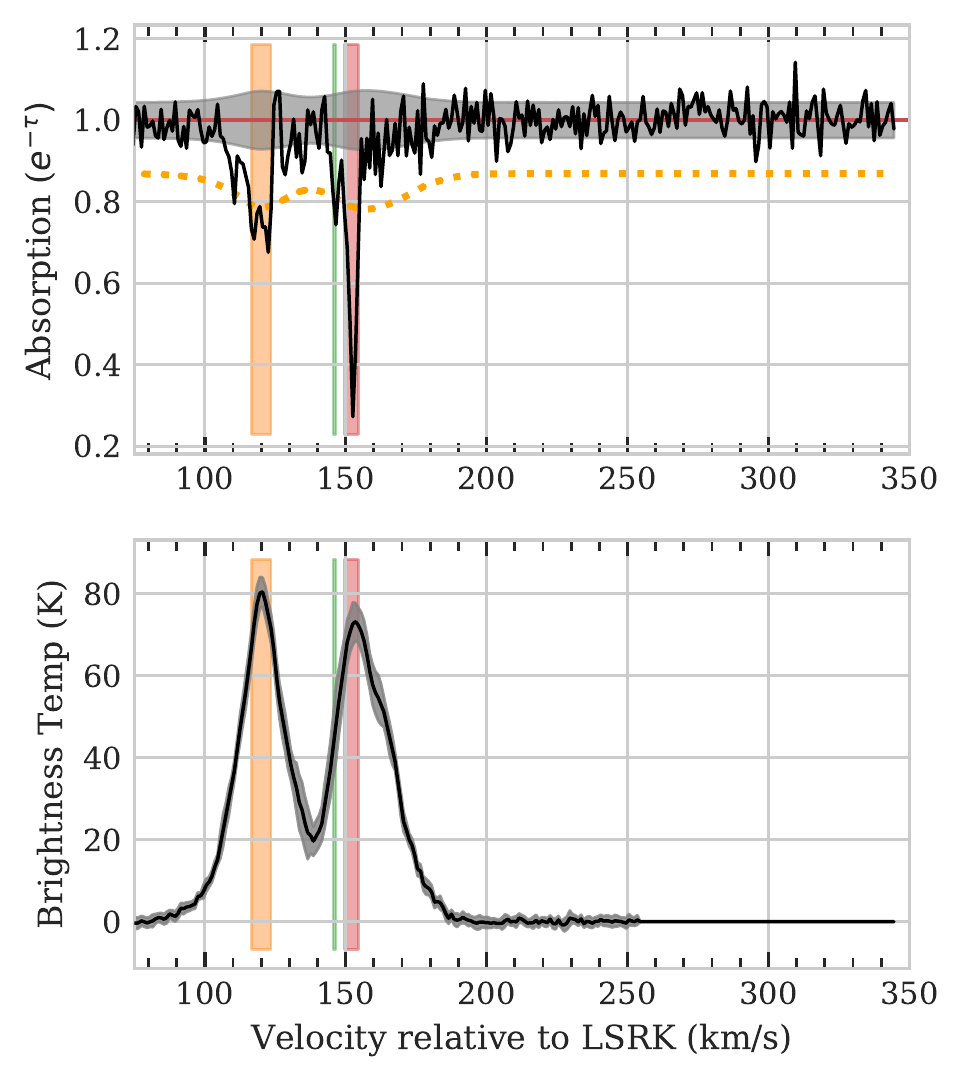}
    \caption{Absorption (top) and emission (bottom) spectra for source J005556$-$722605 in the SMC velocity range. The measured absorption is shown as a black line, the continuum level is shown as a red line, the 1$\sigma$ noise envelope is shaded grey, and the dotted orange line is the 3$\sigma$ absorption level. Regions of detected absorption features are shaded in both the absorption and emission spectra. In the emission plot, the black line shows the brightness temperature, and the 1$\sigma$ uncertainty in the brightness temperature is shown as a grey envelope. Note that the SMC emission cube does not cover the velocity range v$_{\rm LSRK} > 250$ km s$^{-1}$.}
    \label{fig:spectrum}
\end{figure}

An example spectrum for continuum source J005556$-$722605 is shown in Figure \ref{fig:spectrum}.
In the top panel, we can see that the spectral bandpass has a slope of $\approx0$ and is consistent with random noise around a constant continuum level.
The emission has been successfully excluded from the spectrum, with the spectrum above the continuum level consistent with the continuum noise envelope.
In the bottom panel, we show the brightness temperature spectrum from the GASKAP emission cube.
Note that the velocity range of the emission data is restricted to the range of the SMC emission cube.
Overall, the approach of excluding short baselines from the imaging has been successful in excluding emission while maintaining a high signal-to-noise ratio.
The full set of spectra are available in the dataset \citep{Dempsey2021}.

\begin{table*}
    \centering
    \input{spectra_table.tex}
    \caption{Sample of the GASKAP spectrum catalogue. This is a sample of the key fields from the GASKAP HI absorption spectrum catalogue for 14 sources. The full catalogue of all 229 sources is available in the dataset \citep{Dempsey2021}. }
    \label{tab:spectra}
\end{table*}

A subset of sources are described in Table \ref{tab:spectra}, with the full set of sources provided in the dataset. 
For each source we provide 
\begin{enumerate}[label=\alph*)]
    \item ID - the id of the source within the field
    \item Source - name based on the detected location of the source
    \item RA - right ascension of the source
    \item Dec - declination of the source
    \item Peak Flux - peak flux density of the source from the Selavy continuum catalogue
    \item Rating - quality rating of the spectrum (see Sec \ref{sec:specExtract})
    \item $\sigma_{\rm cont}$ - 1$\sigma$ noise level of the spectrum in absorption units. Does not include emission noise
    \item Peak $\tau$ - the maximum $\tau$ value within the SMC velocity range. Where the spectrum is saturated ($e^{-\tau} \le 0$) a minimum limit is specified based on the $1 \sigma$ noise level of the channel in the spectrum
    \item $N_{\rm HI,uncorr}$  - column density towards the continuum source, excluding Milky Way velocities, from the GASS survey under the assumption that the \hi\ is optically thin
    \item $\langle T_{\rm S} \rangle$ - the density-weighted mean spin temperature of the sight-line (see Section \ref{subsec:temperature})
    \item $\mathcal{R}_{\rm HI}$ - column density correction ratio for the sight-line (see Section \ref{subsec:colDensityCorrection}).
\end{enumerate}

\begin{table*}
    \centering
    \input{absorption_table.tex}
    \caption{Sample of the GASKAP absorption feature catalogue. This is a sample of the key fields for those \hi\ absorption features detected in the spectra listed in Table 1. Note that multiple features are detected in some of the spectra, while other spectra have no detectable features.  The full catalogue of all 130 features is available in the dataset \citep{Dempsey2021}.}
    \label{tab:absorption}
\end{table*}

\begin{figure}[ht]
    \centering
    \includegraphics[width=\linewidth]{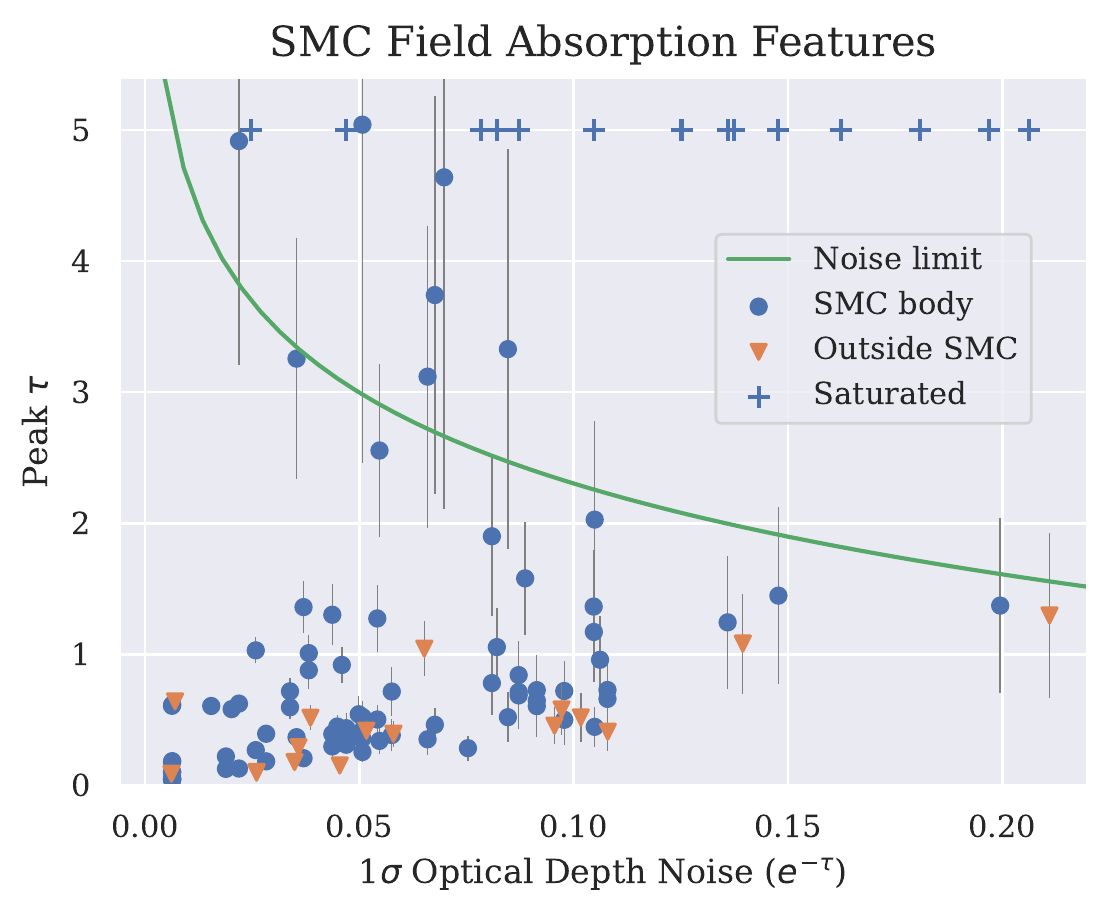}
    \caption{Distribution of the peak optical depth of detected absorption features against the noise in optical depth. Blue dots are non-saturated features in the body of the SMC, orange triangles are non-saturated features outside the body of the SMC.
    Blue plus signs show saturated ($e^{-\tau} < 0$) features in the body of the SMC and the green line shows our $\tau$ sensitivity limit due to noise.}
    \label{fig:opticaldepth}
\end{figure}

\begin{table}
    \centering
    \begin{tabular}{lrrr}
    \hline \hline
    Region & Sources & Detections & Detection Rate \\
    & & & (\%) \\
    \hline
    All & 229 & 65 & 28 \\
    SMC Body & 79 & 49 & 62 \\
    Outside SMC & 150 & 16 & 11 \\
    \hline
    \end{tabular}
    \caption{Detection statistics for different regions}
    \label{tab:detectionrates}
\end{table}

Using the criteria described in Sec.~\ref{sec:specExtract}, we detect absorption features at SMC velocities (v$_{\rm LSRK} \geq 50$ km s$^{-1}$) in 65 (28\%) of the spectra.
Thirty-six of these spectra have multiple features, giving a total of 122 features. 
Note that a full decomposition of these features into components is beyond the scope of this work.
A subset of absorption features is shown in Table \ref{tab:absorption}, with the full set of features provided in the dataset. 
For each feature we provide 
\begin{enumerate}[label=\alph*)]
    \item Source - name of the source based on the location of the source
    \item Feature - name of the feature based on the source name and the minimum velocity of the feature
    \item Min Velocity - the minimum velocity bound of the feature
    \item Max Velocity - the maximum velocity bound of the feature
    \item Width - the number of 1 km s$^{-1}$ channels the feature spans
    \item Peak Absorption - the maximum measured absorption ($1-e^{-\tau}$) of the feature with the uncertainty in absorption spectrum at that velocity, values greater than 1 are saturated
    \item Peak $\tau$ - the maximum $\tau$ value of the feature. Where the spectrum is saturated ($e^{-\tau} \le 0$) a minimum limit is specified based on the noise level in the peak channel
    \item Significance - the highest single-channel significance of the feature as measured in absorption ($e^{-\tau}$)
    \item Equivalent Width - the integral of $\tau$ for the feature.  This will be a lower limit for saturated spectra as we use a minimum limit based on noise for each saturated velocity channel in the same way as peak $\tau$ above.
\end{enumerate}

The distribution of optical depth of these features is shown in Figure~\ref{fig:opticaldepth}.
Sixteen of the features are saturated ($e^{-\tau} < 0$) and 6 features have sufficient noise that they could be saturated (above the noise limit in Fig.~\ref{fig:opticaldepth}).
The $\tau$ values of the 100 remaining features range from 0.04 to as high as 3.3, with a median $\tau = 0.5$.
The trend to have higher $\tau$ values for higher noise spectra is a side-effect of our minimum significance requirement.

Following \cite{Jameson+2019}, we define an arbitrary column density limit for the body of the SMC \footnote{As shown in Fig \ref{fig:columnDensityHist}, the detection rate drops smoothly across this limit so the results are not strongly dependent on the exact value selected.} ($N_{\rm HI,uncorr} > 2 \times 10^{21}$ cm$^{-1}$) and show this as the contour in Figure~\ref{fig:slices}.
Within the body of the SMC, we see a much higher detection rate, as shown in Table \ref{tab:detectionrates}, with absorption features in 62\% of spectra.
Thus 75\% of all spectra with absorption are found in the body of the SMC.
Notably, of the 23 bright ($S_{\rm cont} \ge 50$ mJy) continuum sources within the body of the SMC only one, J003754$-$725157, does not have any detectable absorption features in its spectrum. 
This spectrum has $\tau_{\rm peak} = 0.11\pm0.06$.
So in these higher density regions we see absorption in almost all low noise spectra.
The detection rate also drops with source strength, with only 41\% of the 51 SMC spectra against faint ($S_{\rm cont} < 30$ mJy) continuum sources having detectable absorption.

The 16 features outside the body of the SMC have generally low $\tau$ values within a smaller range (min=0.09, med=0.5 max=1.3) than within the body, and none are saturated.
In contrast to the body of the SMC, only one of the 16 spectra outside the SMC has multiple features.
Of the 71 bright continuum sources outside the SMC, only 7 show detectable absorption features.
As shown in Figure \ref{fig:allsources}, some detections such as J013218$-$715348 and J005715$-$704046 are well away from the SMC. 
Also, as seen in Figure~\ref{fig:slices}, particularly at the extremes of velocity, we detect some shallow absorption features in regions with little emission.
These are discussed in further detail in Section \ref{sec:LowNhAbs}.

\begin{figure*}
    \centering
    \includegraphics[width=\linewidth]{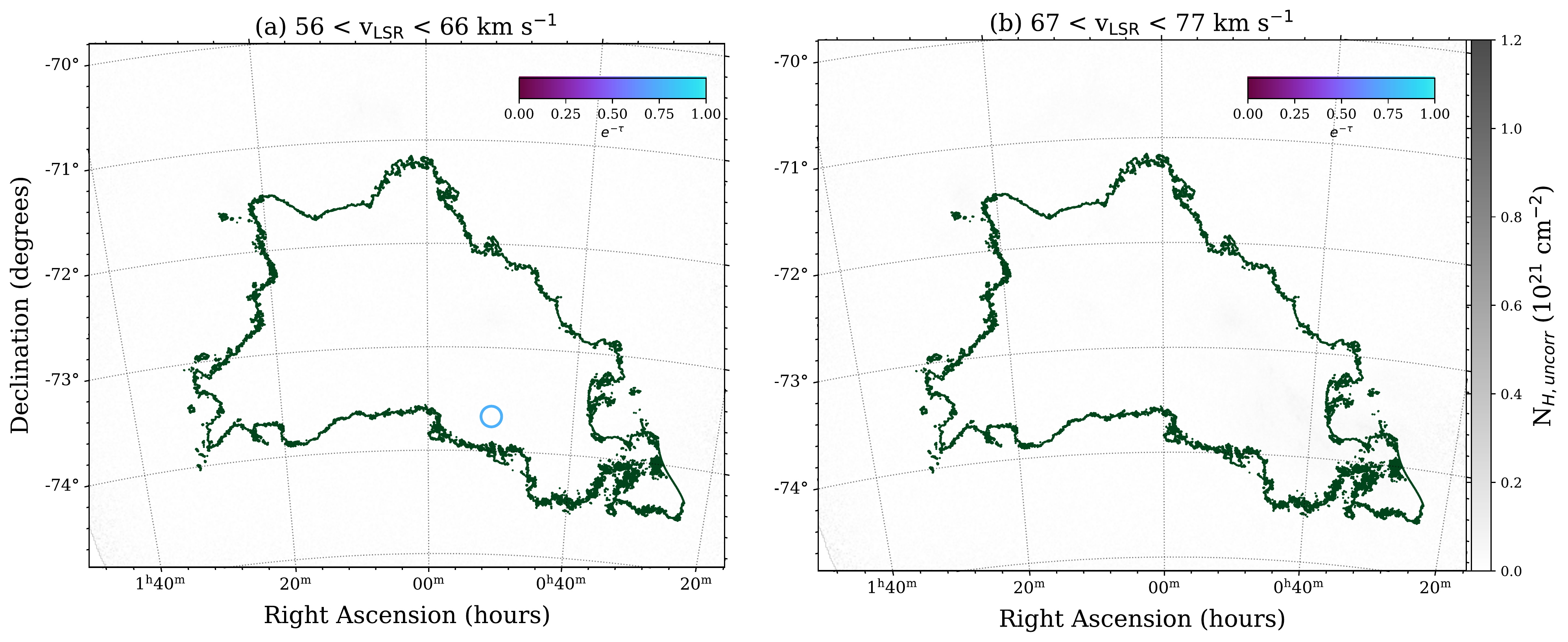}
    \includegraphics[width=\linewidth]{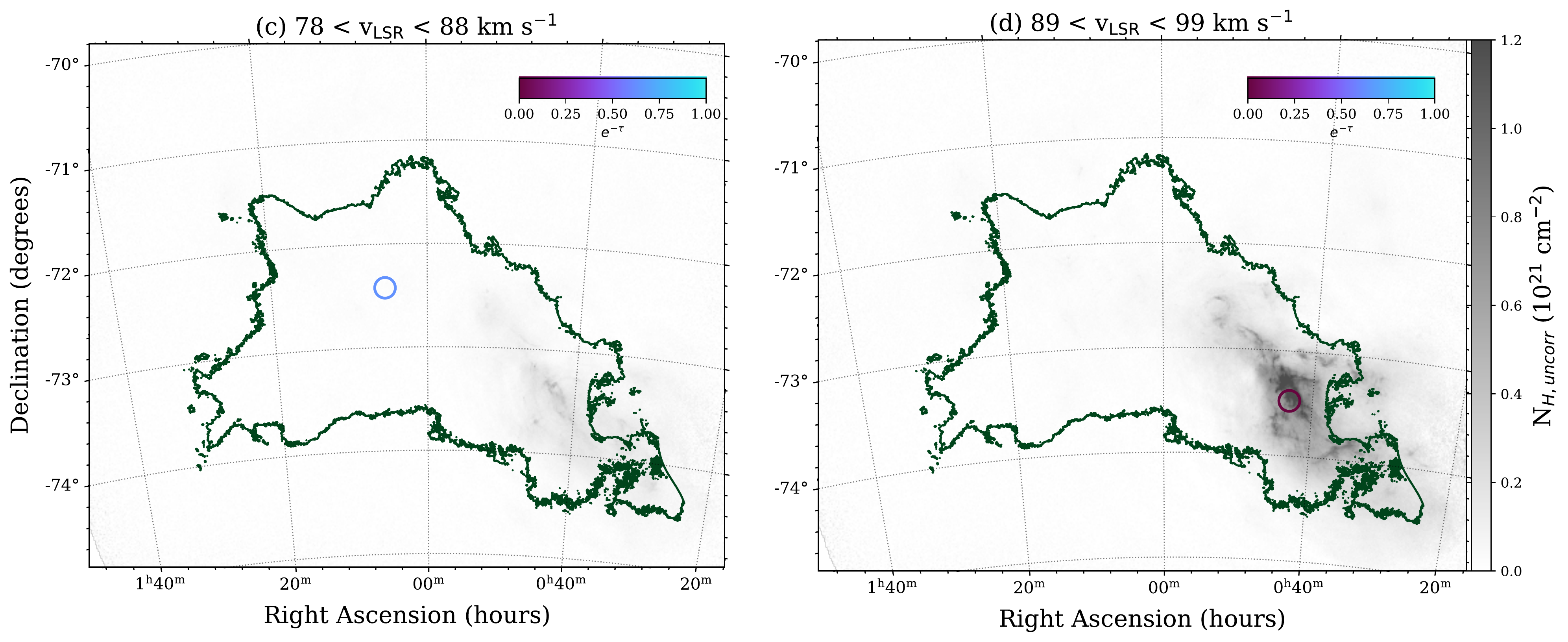}    
    \caption{Locations of the 122 SMC absorption features described in Sec.~\ref{sec:absorption}, plotted against 10 km s$^{-1}$ slices of emission.
    Each absorption feature is shown as a circle centred on its position with the colour reflecting the depth of absorption.
    The scale for the absorption is shown in the top right of each panel, with darker colours indicating deeper absorption and lighter colours shallower absorption. 
    Features are plotted in the velocity slice containing the centre of the absorption feature's velocity range.
    In each slice, the column density for that velocity range, as measured in GASKAP emission \citep{pingel2021}, is shown as a linear grayscale, with the scale shown on the right of each row. 
    The green contour shows the 2$\times 10^{21}$ column density limit for all SMC velocity ranges, as detected in GASKAP emission \citep{pingel2021}, representing the outline of the SMC.}
    \label{fig:slices}
\end{figure*}
\renewcommand{\thefigure}{\arabic{figure} (Cont.)}

\begin{figure*}
    \ContinuedFloat 
    \centering
    \includegraphics[width=\linewidth]{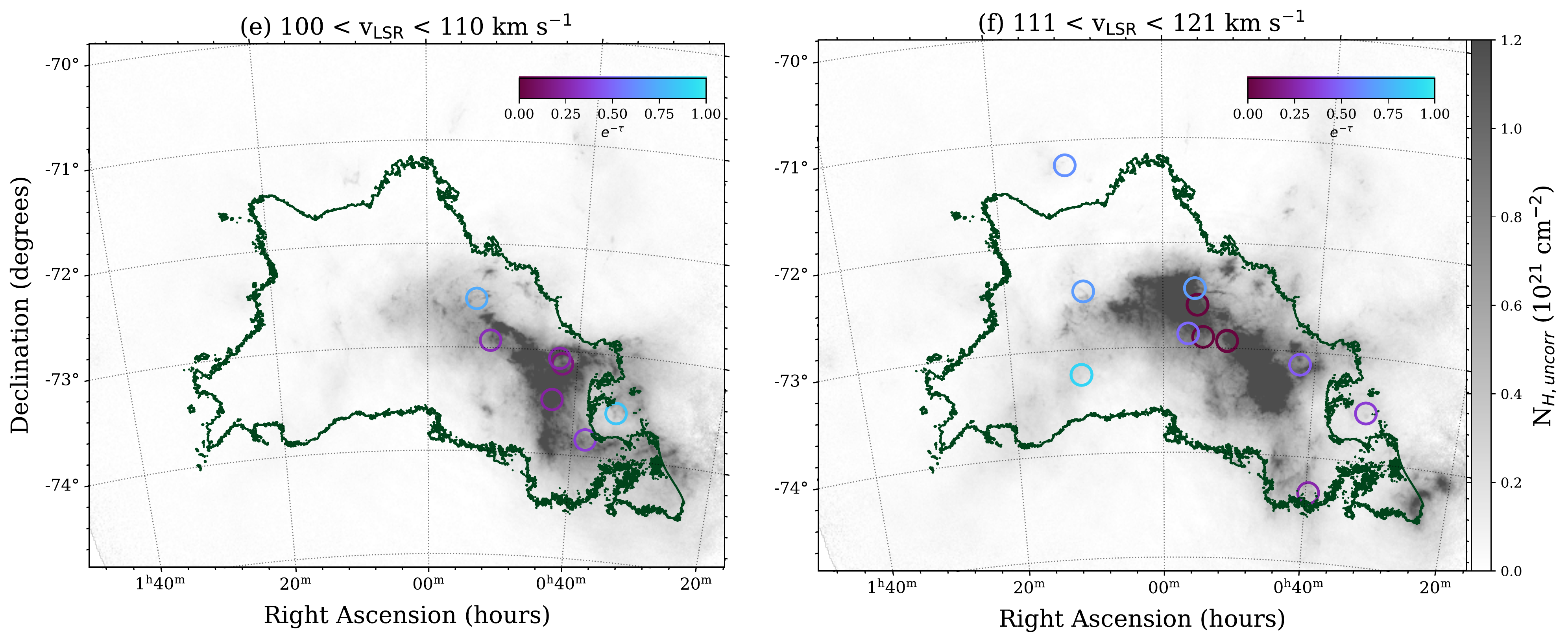}    
    \includegraphics[width=\linewidth]{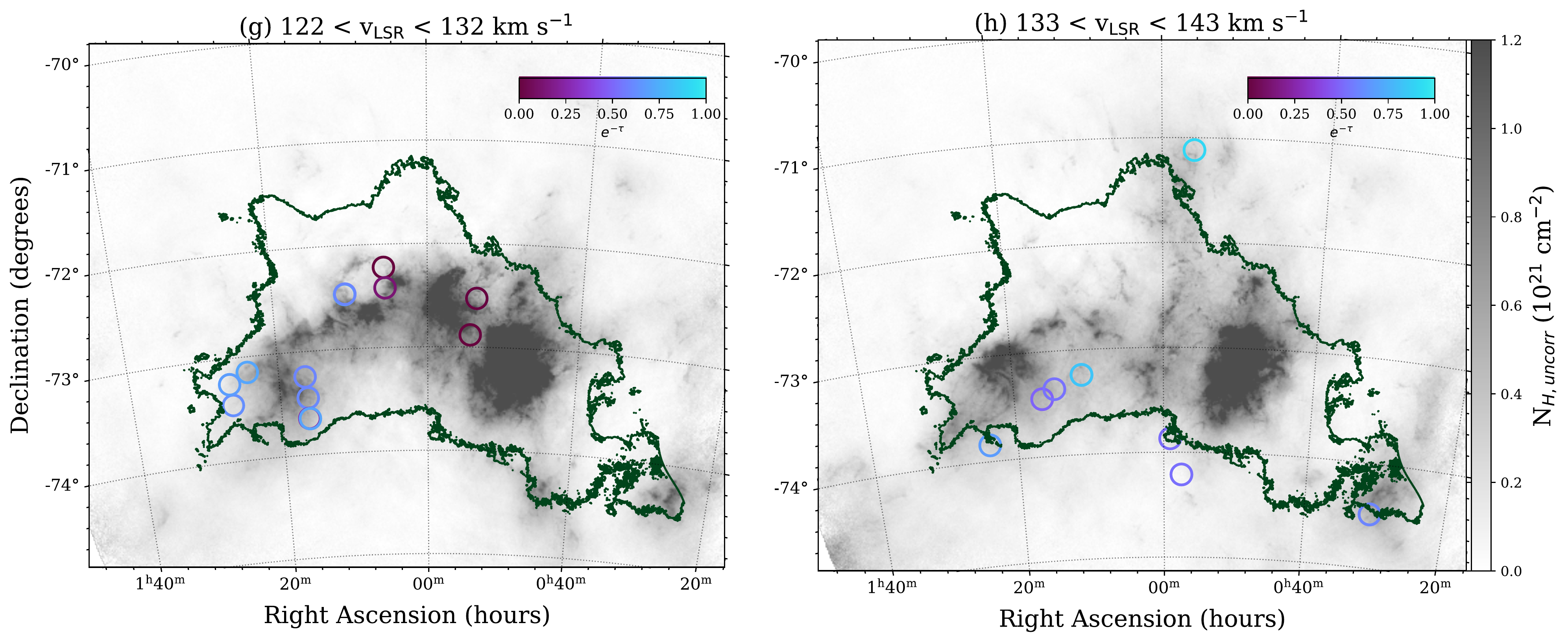}
    \includegraphics[width=\linewidth]{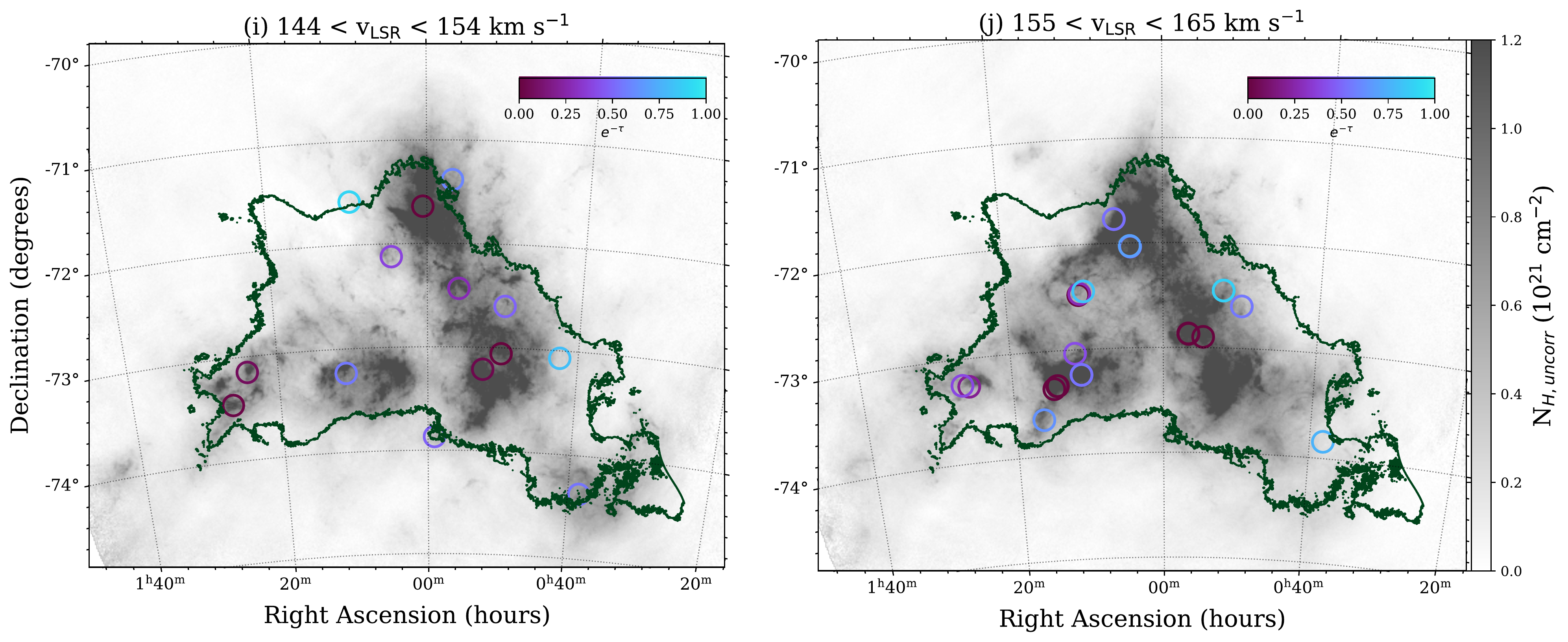}
    \caption{Locations of the SMC absorption features plotted against 10 kms$^{-1}$ slices of emission continued. }
\end{figure*}

\begin{figure*}
    \ContinuedFloat 
    \centering
    \includegraphics[width=\linewidth]{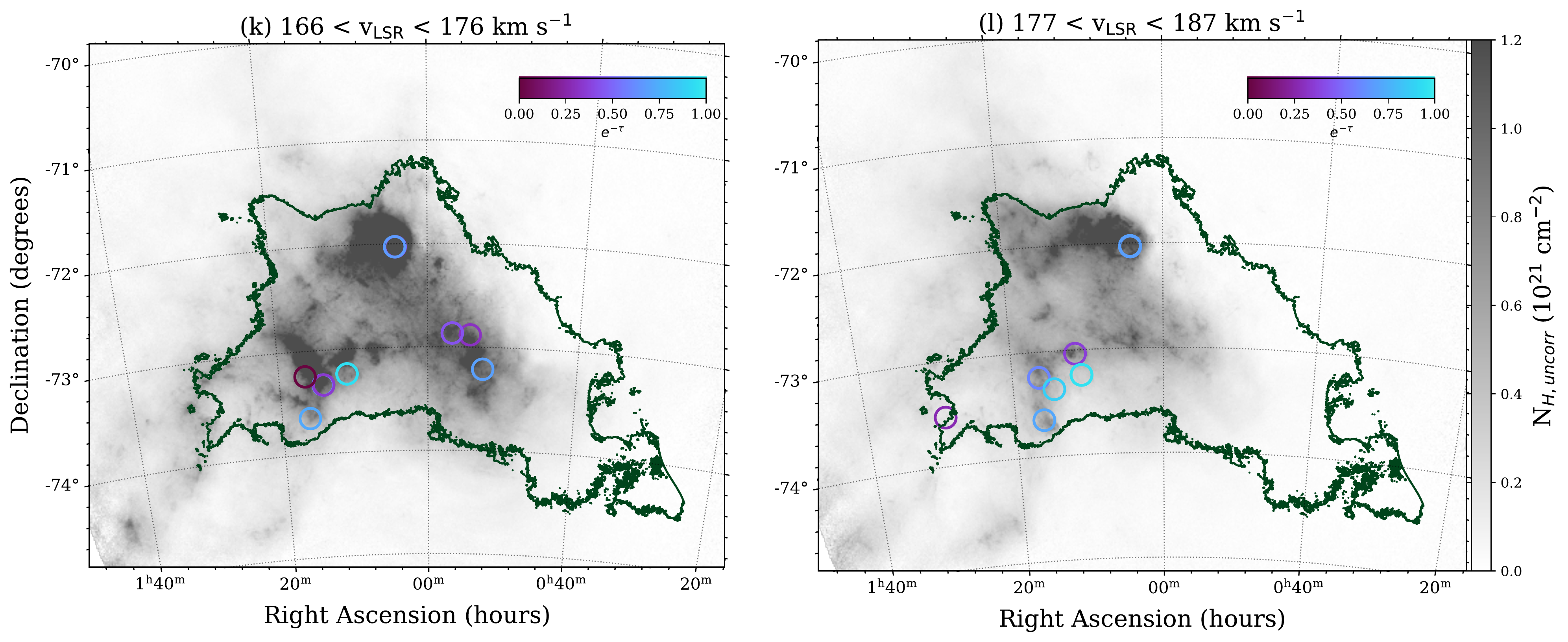}    
    \includegraphics[width=\linewidth]{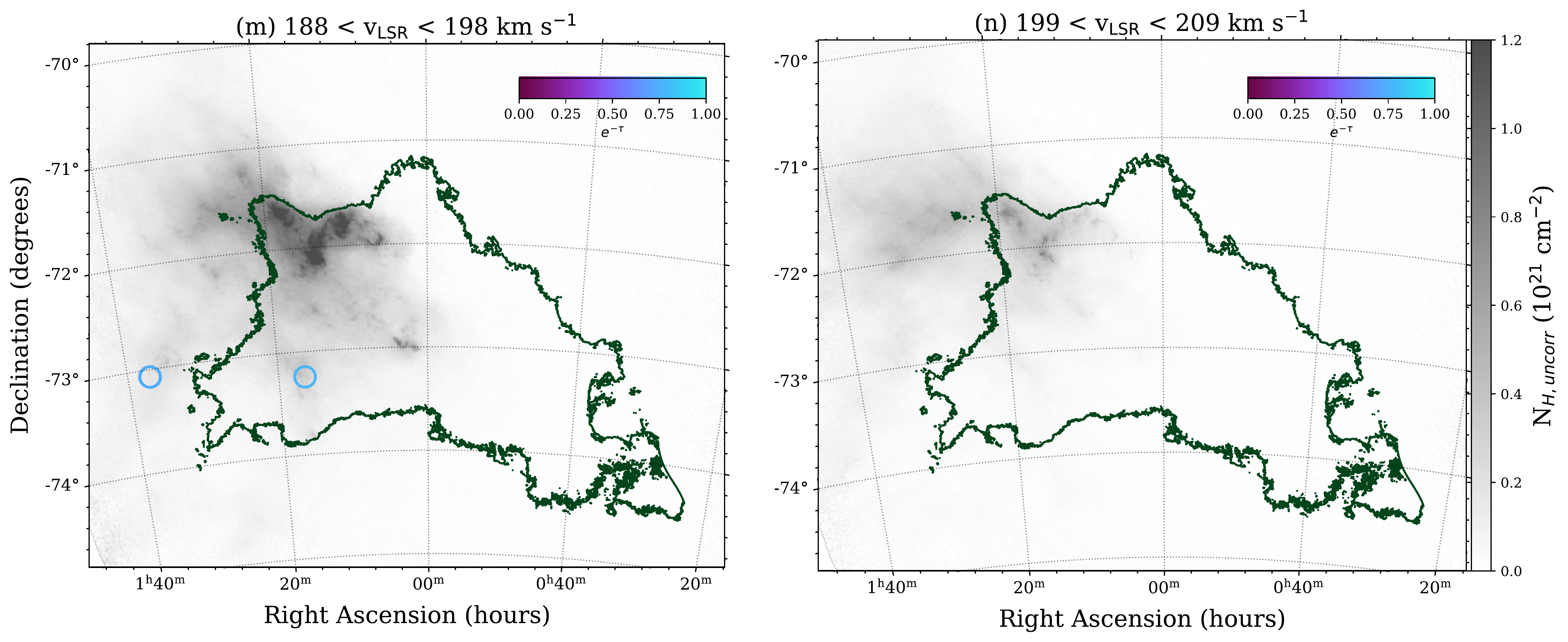}
    \includegraphics[width=\linewidth]{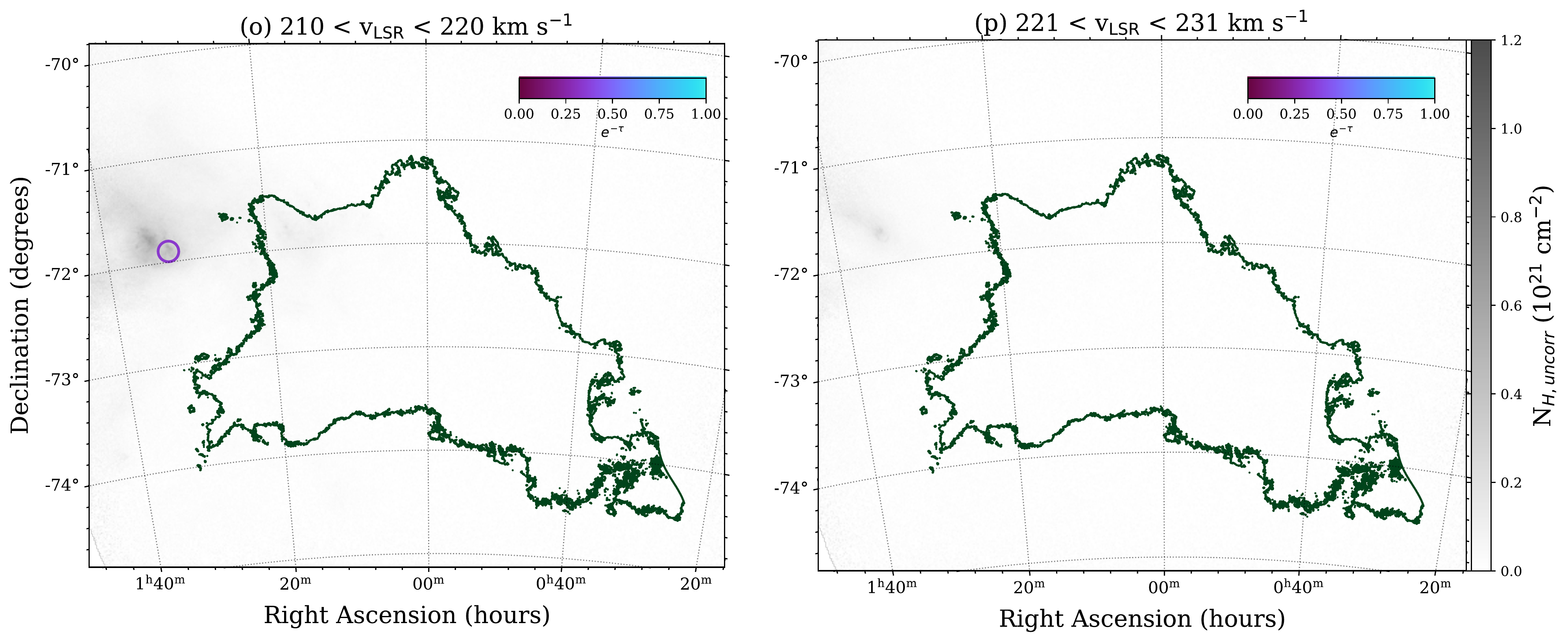}
    \caption{Locations of the SMC absorption features plotted against 10 kms$^{-1}$ slices of emission continued. }
\end{figure*}
\renewcommand{\thefigure}{\arabic{figure}}

\subsection{Absorption in Low Density Regions}
\label{sec:LowNhAbs}

\begin{figure}[ht]
    \centering
    \includegraphics[width=\linewidth]{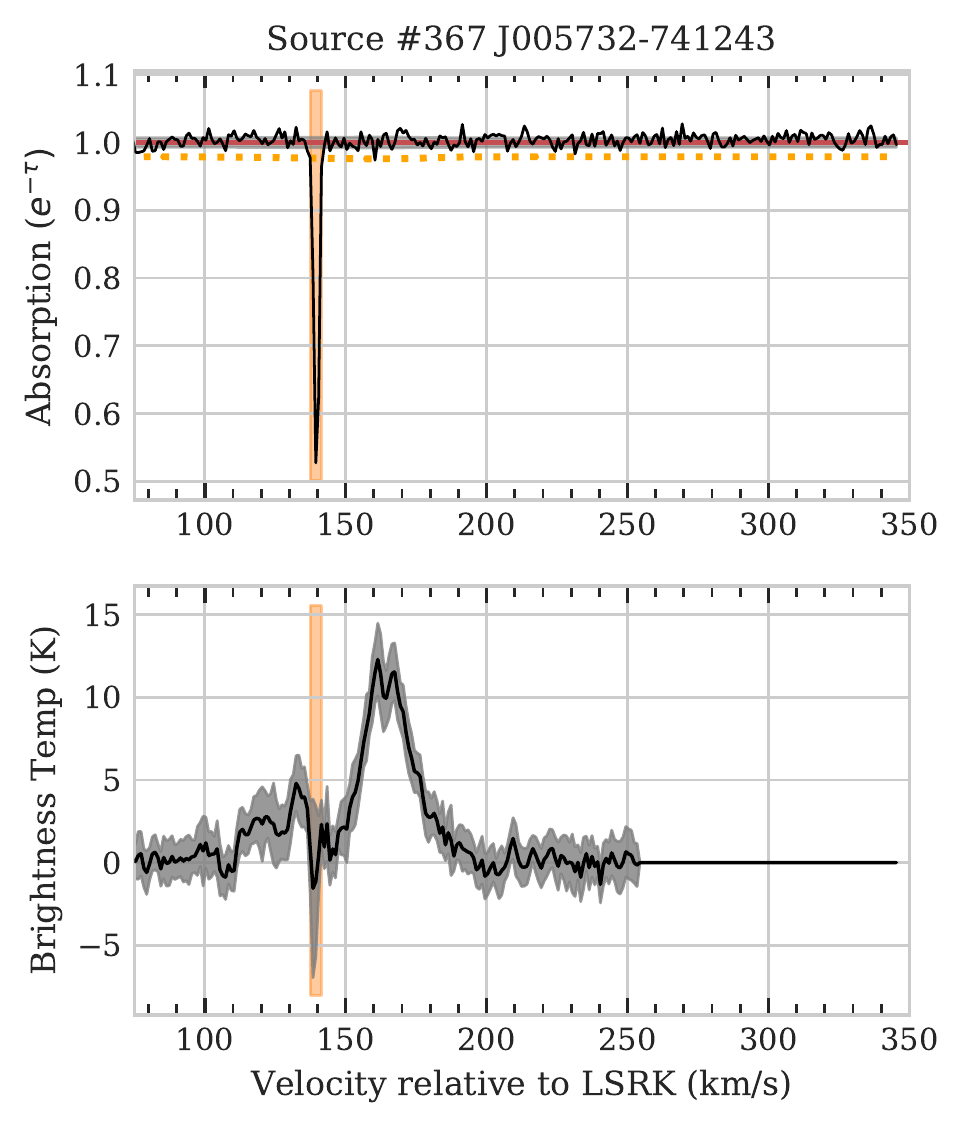}
    \caption{Absorption (top) and emission (bottom) spectra for source J005732$-$741243 in the SMC velocity range. See Figure~\ref{fig:spectrum} for details.}
    \label{fig:spectrum-J005732}
\end{figure}

Six of the detected absorption features are in velocity regions with little emission ($T_{\rm B} < 5$ K), indicating low column density (see Fig.~\ref{fig:slices}).
This is surprising as cold \hi\ is normally associated with a higher density of warm \hi\ \citep[e.g.][]{Kanekar+2011}.
Three of these features are reliable detections.
The first feature, J005732$-$741243\_137 is a very clear detection (significance $\approx$ 61$\sigma$) with deep absorption through low column density (see Fig. \ref{fig:spectrum-J005732}), and is also reported in \cite{Jameson+2019}. 
The feature is also apparent as absorption in the GASKAP emission cube.
Its location, seen in Fig.~\ref{fig:slices}h, is in the lower centre of the image below the SMC body.
The second feature, J005715$-$704046\_255 is outside the velocity range for which we have emission data and thus is not shown in Fig.~\ref{fig:slices}. 
It is located well to the North of the SMC body.
The feature is only two channels wide, but has adjacent channels which also show less significant absorption, making this likely to be a real absorption feature.
A further absorption feature, J011134$-$711414\_114, was detected within a velocity region with $T_{\rm B} = 9\pm4$ K. 
This source is shown in Fig.~\ref{fig:slices}f and is located in the upper left of the image slightly to the North-East of the SMC body.
With two adjacent channels over 3$\sigma$ significance this is a reliable detection.
Of these three features, the latter two are shallow, similar to the low-$N$(\hi) clouds found by \cite{Stanimirovic+2007} in the Milky Way.
They found these type of clouds to be the lower column density population of the CNM but likely to be short-lived due to to their small size and lack of shielding.

The other three features are less firm detections.
In two of these cases (J013134$-$700042\_240 which is not plotted, and J005116$-$734000\_65 in Fig.~\ref{fig:slices}a), the features are marginal detections and consistent with noise.
Notably, the spectrum J013134$-$700042 was flagged as having a potentially underestimated noise level.
Finally, the feature J010524$-$722524\_87 (see Fig.~\ref{fig:slices}c) has a large emission noise spike in the next channel ($v=89$ km s$^{-1}$) to the feature. 
This may indicate that the absorption is a noise artefact also.

\section{Discussion}

\subsection{Column density thresholds and detection rate}
\label{subsec:columnDensity}

Using the GASKAP emission data we can calculate the \hi\ column density for each sightline, under the assumption that the \hi\ gas is optically thin ($\tau \ll 1$) \citep[eq. 3]{Dickey+Lockman90}:
\begin{equation}
    N_{\rm HI,uncorr} = 1.823 \times 10^{18} \int T_{\rm B}(\text{v})~d\text{v}~\text{cm}^{-2}
\end{equation}

We use a Monte-Carlo method to establish the uncertainties in $N_{\rm HI,uncorr}$. 
For each velocity channel of each spectrum we draw 1000 samples of both the optical depth and the brightness temperature.
The samples are randomly drawn from a normal distribution utilising the actual spectrum value as the mean and the 1-$\sigma$ noise envelope as the standard deviation.
We then perform all calculations across all samples, and use the median of the result for each sample as the final value, and the 15th and 85th percentiles as the 1-$\sigma$ uncertainty ranges.

\begin{figure}[ht]
    \centering
    \includegraphics[width=\linewidth]{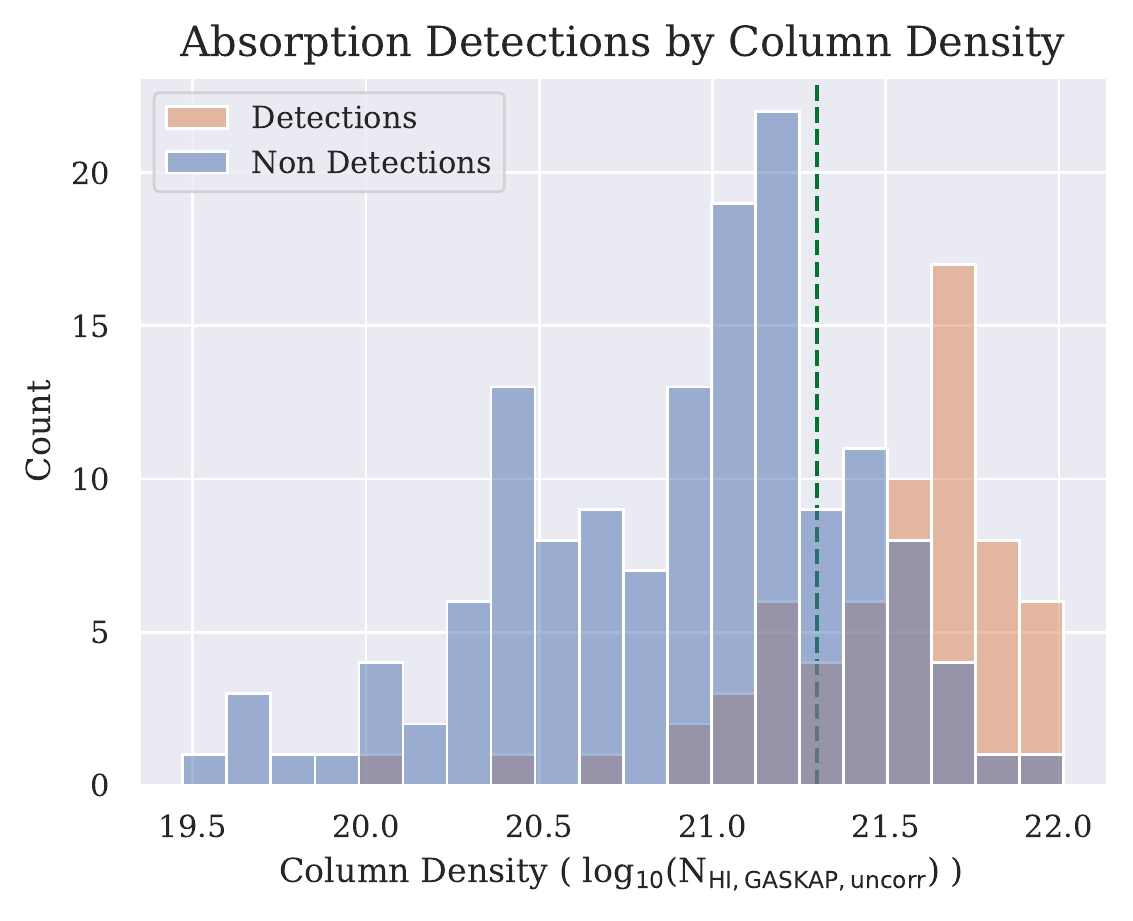}
    \caption{Comparison of the distribution of spectra with and without absorption detections compared with the column density of the lines of sight as measured by GASKAP under the assumption that the \hi\ is optically thin. 
    The blue bars show the count of spectra without detected absorption features in each column density bin, while the orange bars show the count of spectra with absorption features in those bins. The vertical dashed green line marks the $2\times10^{21}$ cm$^{-2}$ column density limit of the SMC body.}
    \label{fig:columnDensityHist}
\end{figure}

With an unbiased sample of absorption in the field we have the opportunity to examine where the cold gas is present.
Figure \ref{fig:columnDensityHist} shows a comparison of the \hi\ column density (uncorrected for optical depth, see Sec.~\ref{subsec:colDensityCorrection}) for both spectra with and without absorption detections.
We can see that we have some detections in regions with column densities as low as $1.25 \times 10^{20}$ cm$^{-2}$.
However, it is not until $2.5 \times 10^{21}$ cm$^{-2}$ that we have a high rate of detection.
Most absorption detections are in sight-lines with a column density of $3 \times 10^{21}$ cm$^{-2}$ or greater.
From $6 \times 10^{21}$ cm$^{-2}$ almost all sight-lines, even for faint continuum sources, show absorption.

\cite{Kanekar+2011} presented a relationship between the observables of uncorrected column density and the integral of $\tau$ (the equivalent width) in Milky Way gas clouds.
The following equation (their Equation~3) models the presence of cold gas within warm \hi\ envelopes, which shield the cold \hi\ from surrounding hot gas and radiation.
\begin{equation}
    N_{\rm HI,uncorr} = N_0 e^{-\tau^{\prime}_{c}} + N_{\infty} (1-e^{-\tau^{\prime}_{c}})
    \label{eq:kbr11}
\end{equation}
Here, $N_0$ is the threshold column density for the formation of cold gas, $N_{\infty}$ is the column density at which the cold gas saturates ($\tau \to \infty$) and $\tau^{\prime}_{c}$ is the effective opacity, or the mean observed optical depth over the velocity range ($\Delta V$).
They further define a lower and upper limit using Eq.~\ref{eq:kbr11} on the expected range in which absorption forms in the Milky Way.
The lower limit has $N_0 = 10^{20}$ cm$^{-2}$, $N_{\infty} = 5.0 \times 10^{21}$ cm$^{-2}$ and $\Delta V = 20$ km s$^{-1}$.
The upper limit has $N_0 = 2 \times 10^{20}$ cm$^{-2}$, $N_{\infty} = 10^{22}$ cm$^{-2}$ and $\Delta V = 10$ km s$^{-1}$.

\begin{figure}[ht]
    \centering
    \includegraphics[width=\linewidth]{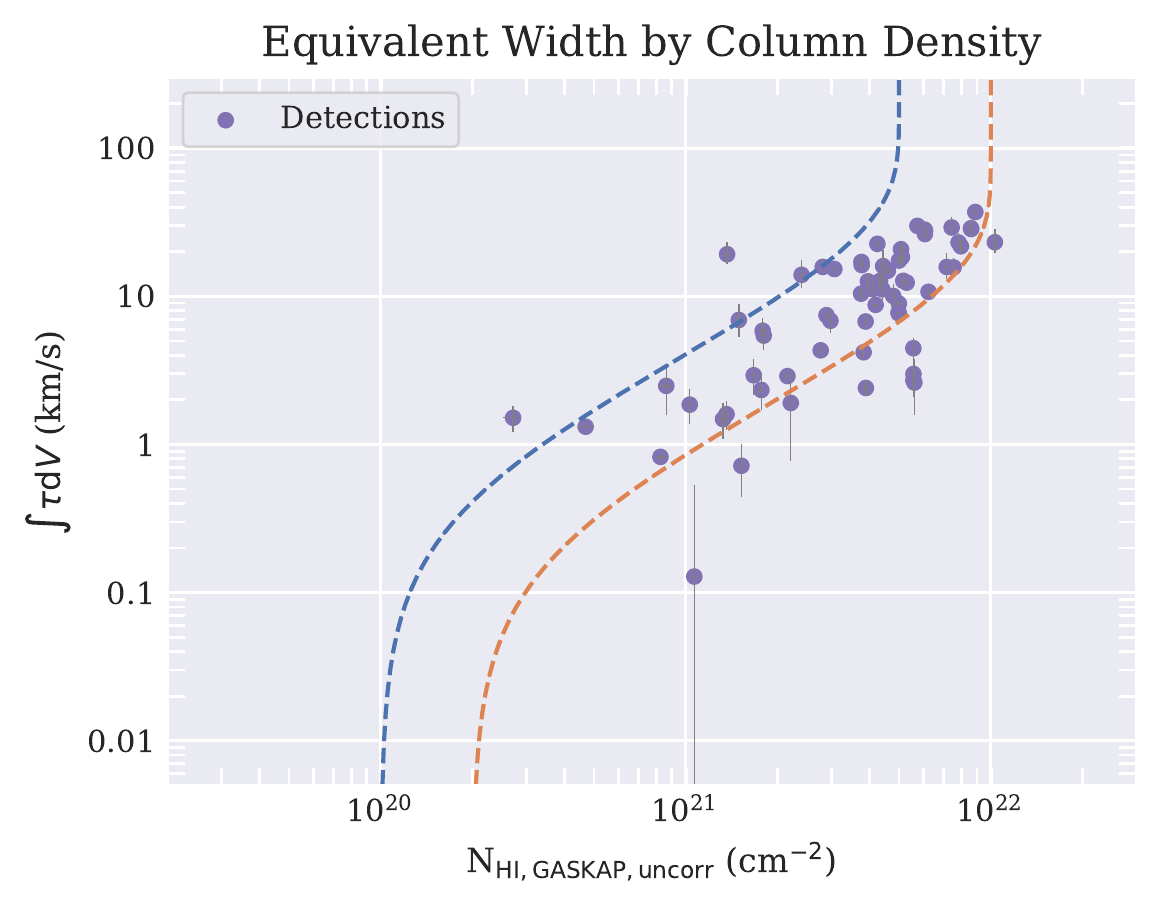}
    \caption{Comparison of uncorrected column density with integrated optical depth (Equivalent Width). The curved lines are the lower (blue) and upper (red) limits defined in \cite{Kanekar+2011} section 3.}
    \label{fig:intOpticalDepth}
\end{figure}

Comparing our detections against the \cite{Kanekar+2011} limits (see Figure~\ref{fig:intOpticalDepth}), we see outliers on both sides of the envelope, despite most sources lying within the limits.
The two sources to the left of the lower limit hint at cold gas formation below the column density limit seen in the Milky Way.
In the current data we do not have the sensitivity to probe this lower column density regime in detail.
To the right of the upper limit we find eight sources.
Four of these sources have low equivalent width noise levels and thus support a higher limit for the condensation of dense \hi\ into molecular H$_2$ saturation in the SMC. 
This higher limit reflects the findings of both \cite{Bialy+16} and \cite{Krumholz+2009} that higher column density is required at low metallicities for the formation of molecular gas, and of \cite{Bolatto+11} that the SMC has a very low fraction of molecular gas as compared to its atomic gas content.
With the addition of future GASKAP observations of the SMC we should be able to test these bounds more rigorously.

\subsection{Corrected column density for the SMC field}
\label{subsec:colDensityCorrection}

The column density measurements from emission data are made under the assumption that the gas is optically thin ($\tau \ll 1$). 
However, if the \hi\ is not optically thin we need to account for the absorption to get an accurate estimation of the column density.
For the sight-lines we have observed, we can correct these measurements for the actual optical depth of the gas.
Using the assumption that the gas is isothermal, that is each velocity channel of gas only has a single temperature, we can calculate the corrected column density using the formula (\citealt[eq 5]{DB82} and \citealt[eq 8]{Chengalur+2013}):
\begin{equation}
N_{\rm HI,corr,iso} = 1.823 \times 10^{18} \int \frac{T_{\rm B}({\rm v}) \times \tau({\rm v})}{1-e^{-\tau({\rm v})}}\,d{\rm v}\,{\rm cm}^{-2}
\end{equation}
where $T_{\rm B}({\rm v})$ and $\tau({\rm v})$ are the brightness temperature and absorption respectively for a velocity channel.
The column density correction ratio is then  
\begin{equation}
    \mathcal{R}_{\rm HI} = \frac{N_{\rm HI,corr,iso}}{N_{\rm HI,uncorr,iso}},
\end{equation}
We use a Monte-Carlo method to establish the uncertainties in $N_{\rm H,corr,iso}$ and $\mathcal{R}_{\rm HI}$, as described in Sec.~\ref{subsec:columnDensity}.

\begin{figure}[ht]
    \centering
    \includegraphics[width=\linewidth]{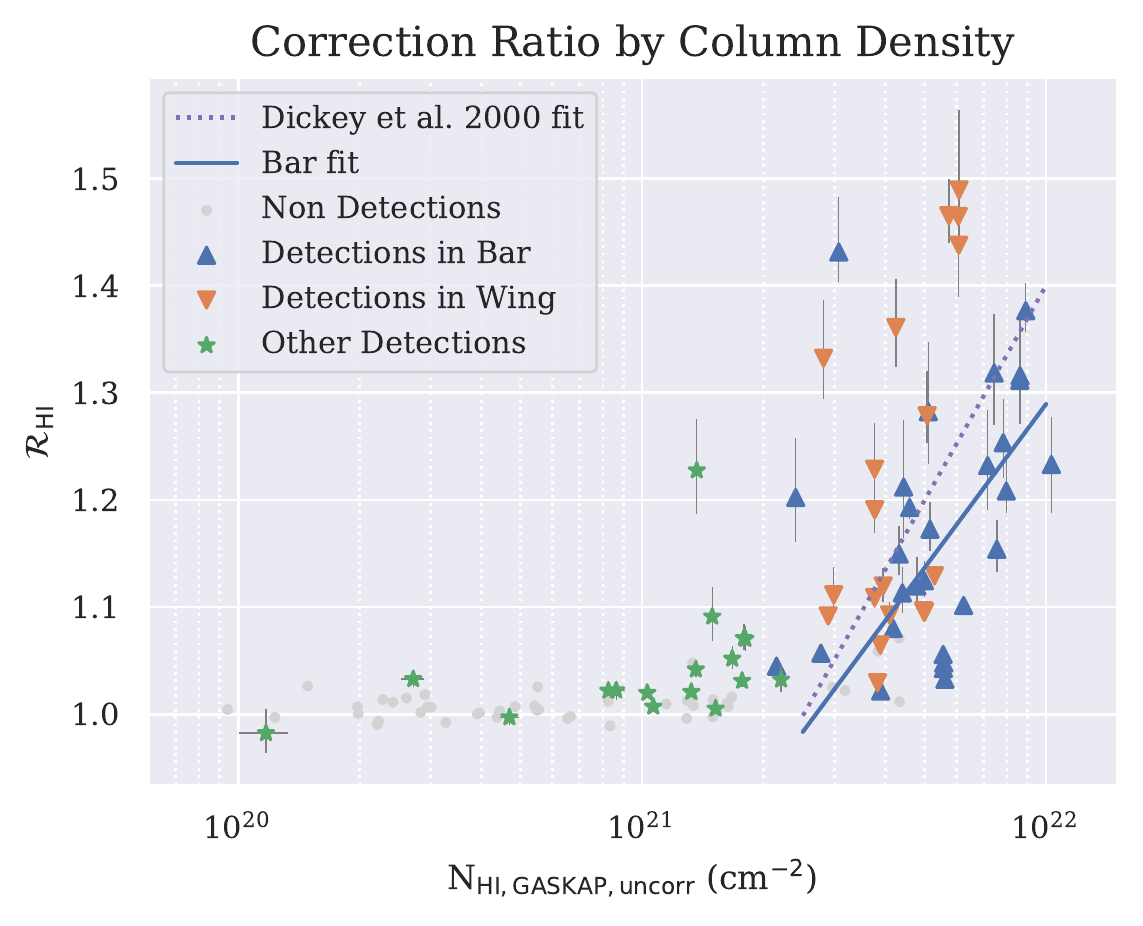}
    \caption{Comparison of uncorrected column density with the correction factor due to optical depth measured for sight-lines with detected absorption.}
    \label{fig:columnDensityCorrection}
\end{figure}

There is a large spread in the correction factors as compared to uncorrected column density, as shown in Fig.~\ref{fig:columnDensityCorrection}. 
Strikingly, the range of correction factors is very different between the wing and the bar of the SMC.
The bar shows generally low correction factors with an increase in correction factor with increasing column density. The exception is the sight-line towards J010029$-$713826, which shows both a deep and a wide absorption component in a region with relatively low column density for the bar region.
In contrast, the wing shows a great variety of correction factors unrelated to column density.
This likely reflects that the bar has a longer line of sight than the wing, and so multiple features would be blurred.
In the shorter line of sight through the wing, we are more likely to pick out individual features such as diffuse tidal structures and regions of intense star forming.
The cluster of four sight-lines with the highest column density and highest correction factor in the wing are all close to NGC 460 and NGC 465. 
The other two sight-lines with $\mathcal{R}_{\rm HI} > 1.3$ are towards the SE end of the wing, another region of active star formation.
The detections outside the wing and the bar, all at lower column densities, show small correction factors as expected.
The one exception is the sight-line towards J012924$-$733153, in which the correction factor is boosted by noise ($\sigma_{cont} = 0.211$) as well as the single narrow detection and a potential shallow, wide component.
For comparison we also show the correction factors of low-noise ($\sigma_{cont} < 0.1$) sight-lines where no absorption was detected. 
These are mostly at lower column densities and have minimal correction, as expected.

Overall, the correction factors range from $\mathcal{R}_{\rm HI} \approx 1$ at the low column densities, to $\mathcal{R}_{\rm HI} = 1.49$ in the higher density regions of the wing, close to known star formation regions.
\cite{Dickey+00} found a correction factor relationship with column density for the SMC of $\mathcal{R}_{\rm HI} \simeq 1 + 0.667(\text{log}_{10}N - 21.4)$. 
With our broader view of the SMC optical depth, we find a slightly shallower linear rise in correction factor with column density for just the SMC bar of $\mathcal{R}_{\rm HI} \simeq 1 + 0.51(\text{log}_{10}N - 21.43)$.
The wide range of correction factors in the SMC wing means that a single line cannot be fit to that region.
The wide range for the SMC wing is similar to the results of \cite{Nguyen+2019} who found large regional variation in $\mathcal{R}_{\rm HI}$ around five giant molecular clouds.

The shallower relationship of column density and correction factor than found in \cite{Dickey+00} has two drivers.
Firstly, as noted in Sec.~\ref{subsec:coldGasFraction}, we detect a higher proportion of shallower absorption features due to the unbiased selection of continuum sources.
Secondly, the smaller emission beam size will reduce beam smearing of small-scale dense structure, so in these cases we will measure a higher uncorrected column density. 
One of the great advantages GASKAP has over past absorption studies is that we have a very closely matched beam size between the emission (30x30 arcsec$^2$) and absorption (15x15 arcsec$^2$).
This means both that our emission sample is taken from as close as possible to the target continuum source and that the scale of structures detected by the emission and absorption are similar.

Comparing the cumulative distribution of correction factors between the field outside the SMC and the body of the SMC, we find that the body has far higher correction factors (see Fig.~\ref{fig:rhicdf}, top).
The higher correction factors reflect the column density threshold apparent at $N_{\rm HI,uncorr} < 2 \times 10^{21}$ cm$^{-2}$ in Fig.~\ref{fig:columnDensityCorrection}, below which there are much lower correction factors.
The uncertainties in this top panel are generated from the $\mathcal{R}_{\rm HI}$ uncertainties for each source. 

\subsection{Temperature distribution}
\label{subsec:temperature}

To measure the density-weighted mean spin temperature we can compare the integrals of brightness temperature and optical depth using the following formula \citep[Eq. 4]{Dickey+00}:
\begin{equation}
    \langle T_{\rm S} \rangle = \frac{\int T_{\rm B}(\text{v})~d\text{v}}{\int 1-e^{-\tau(\text{v})}~d\text{v} } = \int \frac{n(\text{s})}{[n(\text{s})/T_{\rm S}(\text{s})]} d\text{s} ,
\end{equation}
where $n(\text{s})$ is the line-of-sight volume density.
We use a Monte-Carlo method to establish the uncertainties in $ \langle T_{\rm S} \rangle$, as described in Sec.~\ref{subsec:columnDensity}.

Following \cite{Dickey+00} we measure the integrals across the velocity range where $T_B \geq 3$ K so as to reduce the noise in the integrals.
Note that in some cases the integral range will contain regions where the emission level dips below this threshold (e.g. where the sight-line has a bimodal emission distribution).
In those cases we still include the emission from this inner velocity region in the integral.

\begin{figure}[ht]
    \centering
    \includegraphics[width=\linewidth]{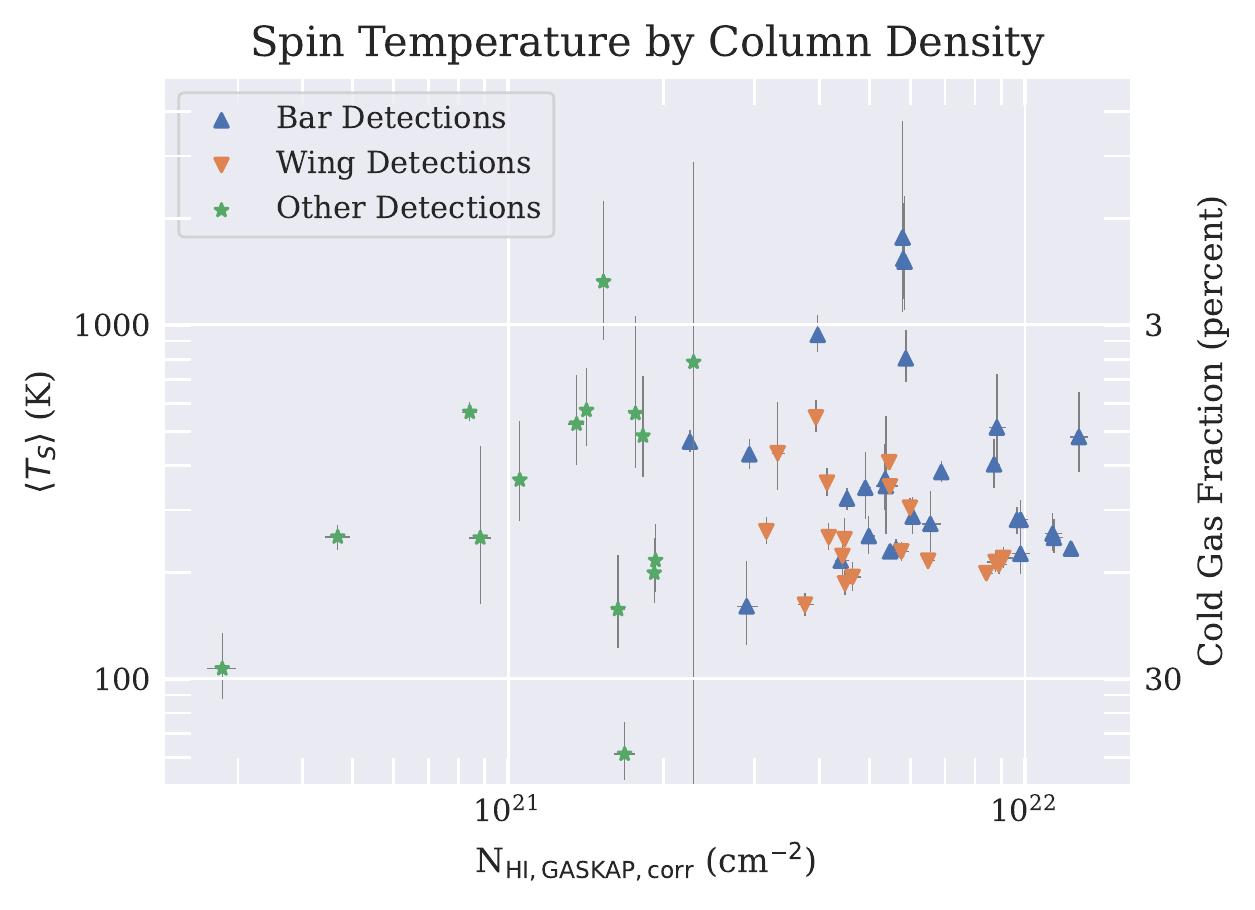}
    \includegraphics[width=\linewidth]{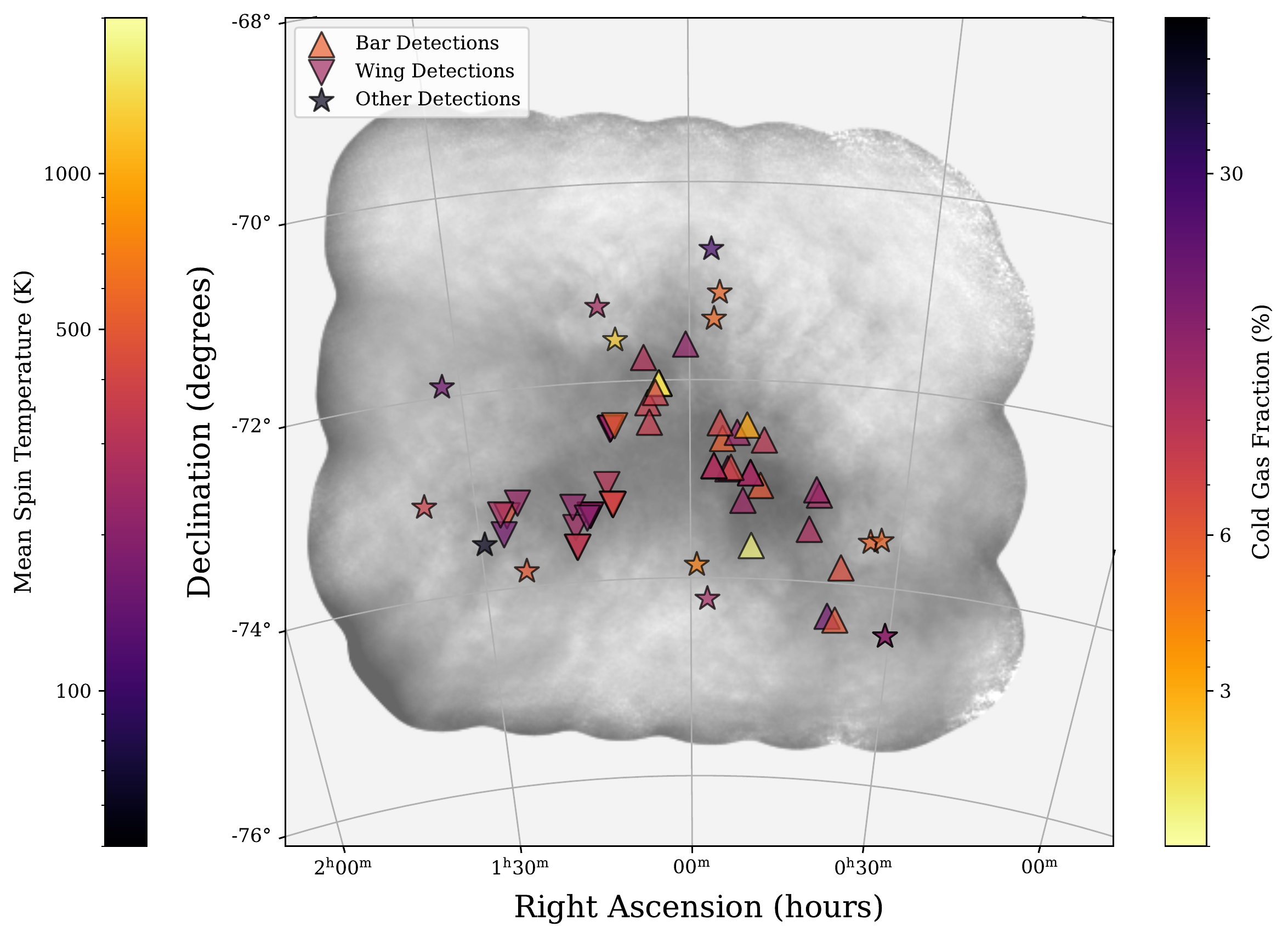}
    \caption{(Top) Comparison of column density with density weighted mean spin temperatures for sight-lines with detected absorption. The column density is calculated using GASKAP emission data and corrected for absorption, see Sec.~\ref{subsec:colDensityCorrection}. (Bottom) Distribution of the mean spin temperatures against the SMC \hi\ column density map from GASKAP \citep{pingel2021}.}
    \label{fig:spinTemp}
\end{figure}

Within the SMC ($N_{\rm HI,GASKAP,uncorr} \geq 2\times10^{21}$ cm$^{-2}$) we find an inverse-noise weighted mean spin temperature of
$\langle T_{\rm S} \rangle = 245\pm2$ K.
This is lower than the 350~K found by  \cite{Dickey+00}.
\cite{Jameson+2019} found an arithmetic mean spin temperature of $\langle T_{\rm S} \rangle = 117.2 \pm 101.7$ K, reflecting their targeting of cold gas.
Outside the SMC we find $\langle T_{\rm S} \rangle = 156^{+9}_{-7}$ K.
The spread of mean spin temperatures against corrected column density for all sight-lines with detected absorption is shown in Fig.~\ref{fig:spinTemp}.

The distribution both by temperature and by corrected column density also varies by region, as shown in Fig.~\ref{fig:spinTemp}.
As was done in \cite{McClure-Griffiths+18}, we split the SMC into two regions, the bar running from N to SW and the wing from N to SE (see Fig.~\ref{fig:spinTemp} bottom). 
The bar has a large spread in both column density and in mean spin temperature. It contains the highest densities and the highest temperatures of sight-lines in the SMC.
However, there is no significant trend in temperature as column density increases in the bar.
The wing has a lower column density than much of the bar and both column density and mean spin temperatures are more tightly clustered than the bar.
Despite the greatly varying environment, the distribution of spin temperatures in the wing is remarkably flat ($163 < \langle T_{\rm S,wing} \rangle < 546$ K) with no strong trend with column density.
In the lower column density regions outside the SMC, there is a much wider spread of spin temperatures and again there is no significant trend of temperature with column density.

In two cases (J013134$-$700042 and J013701$-$730415) our spin temperatures are unconstrained. If there is sufficient emission pollution of the absorption spectrum to drive the integral of $1-e^{-\tau}$ negative, then the calculated mean spin temperature will be negative.
As noted in section \ref{sec:LowNhAbs},  the absorption in J013134$-$700042 is likely a noise artefact and this noise also results in narrow spikes of emission in the absorption spectrum.
For J013701$-$730415, while the absorption is a clear detection, there are also more random emission spikes than expected in the spectrum.

\subsection{Fraction of Cold Gas}
\label{subsec:coldGasFraction}

By assuming a temperature of the cool gas ($T_{\rm c}$) we can estimate the fraction of cool gas in each sight-line as follows \citep[Eq. 7]{Dickey+00}:
\begin{equation}
    f_{\rm c} \equiv \frac{N_{\rm c}}{N_{w} + N_{c}} \simeq \frac{T_{\rm c}}{\langle T_{\rm s} \rangle}  ,
\end{equation}
where $N_{\rm c}$ is the column density of the CNM and $N_{\rm w}$ is the WNM column density.

\begin{figure}[ht]
    \centering
    \includegraphics[width=\linewidth]{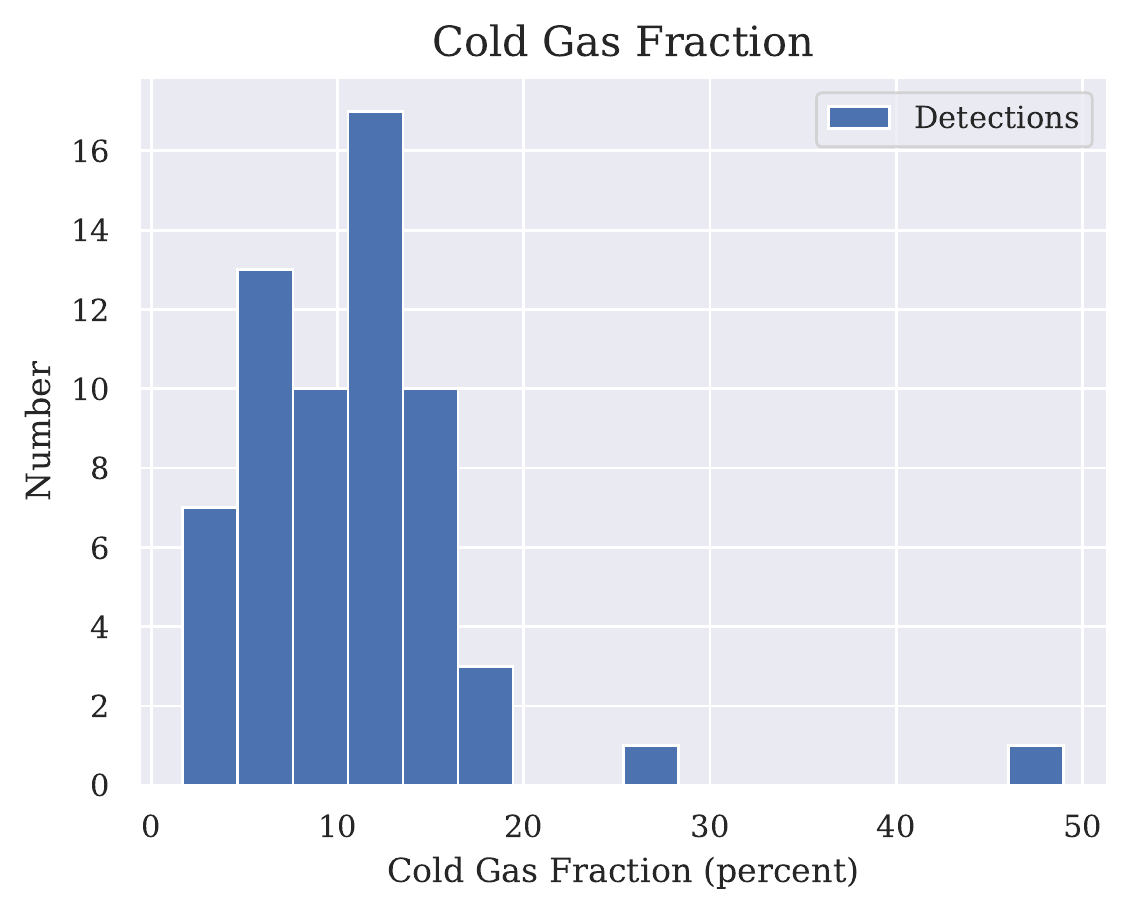}
    \caption{Histogram of the distribution of cold gas fractions for sight-lines with detected absorption.}
    \label{fig:coldGasFraction}
\end{figure}

Following \cite{Jameson+2019}, we assume $T_{\rm c} = 30$ K.
This gives a median cold gas fraction across the field of $f_{\rm c} \approx 11\%$ with the distribution as shown in Fig.~\ref{fig:coldGasFraction}.
The distribution within and outside the body of the SMC is not significantly different.
However, this is significantly lower than the $f_{\rm c} \approx 20\%$ found by \cite{Jameson+2019}.
\cite{Dickey+00} reported a median cold gas fraction of $13\%$ for $T_{\rm c} = 55$ K, which is equivalent to $f_{\rm c} \approx 7\%$ for $T_{\rm c} = 30$ K.
Our result sits between these two earlier SMC results, similarly to the mean spin temperature results in Sec.~\ref{subsec:temperature}, as would be expected.

The higher mean spin temperatures and lower cold gas fractions than those found in the targeted \cite{Jameson+2019} survey are driven by our detection of shallower absorption in the unbiased sight-lines of the present study.
We can expect that with the addition of future GASKAP observations of the SMC we will see further shallow absorption detections and thus an even lower median cold gas fraction.
The key results of this survey and the earlier SMC surveys are shown in Table \ref{tab:surveyResults}.

\begin{table}
    \centering
    \begin{tabular}{lrr}
    \hline \hline
    Survey & $\langle T_{\rm S} \rangle$ \footnote{See Sec. \ref{subsec:temperature} for details of uncertainties in $\langle T_{\rm S} \rangle$} & $f_{\rm c}$ (\%) \\
     & (K) & ($T_{\rm c}=30$ K)   \\
    \hline
    \cite{Dickey+00} & 350 & 7    \\
    \cite{Jameson+2019} & $117.2\pm 101.7$ & 20    \\
    This work & $245\pm2$ & 11 \\
    \hline
    \end{tabular}
    \caption{Comparison of the results of SMC absorption surveys}
    \label{tab:surveyResults}
\end{table}

\subsection{Comparison with the Milky Way}

\begin{figure}[ht]
    \centering
    \includegraphics[width=\linewidth]{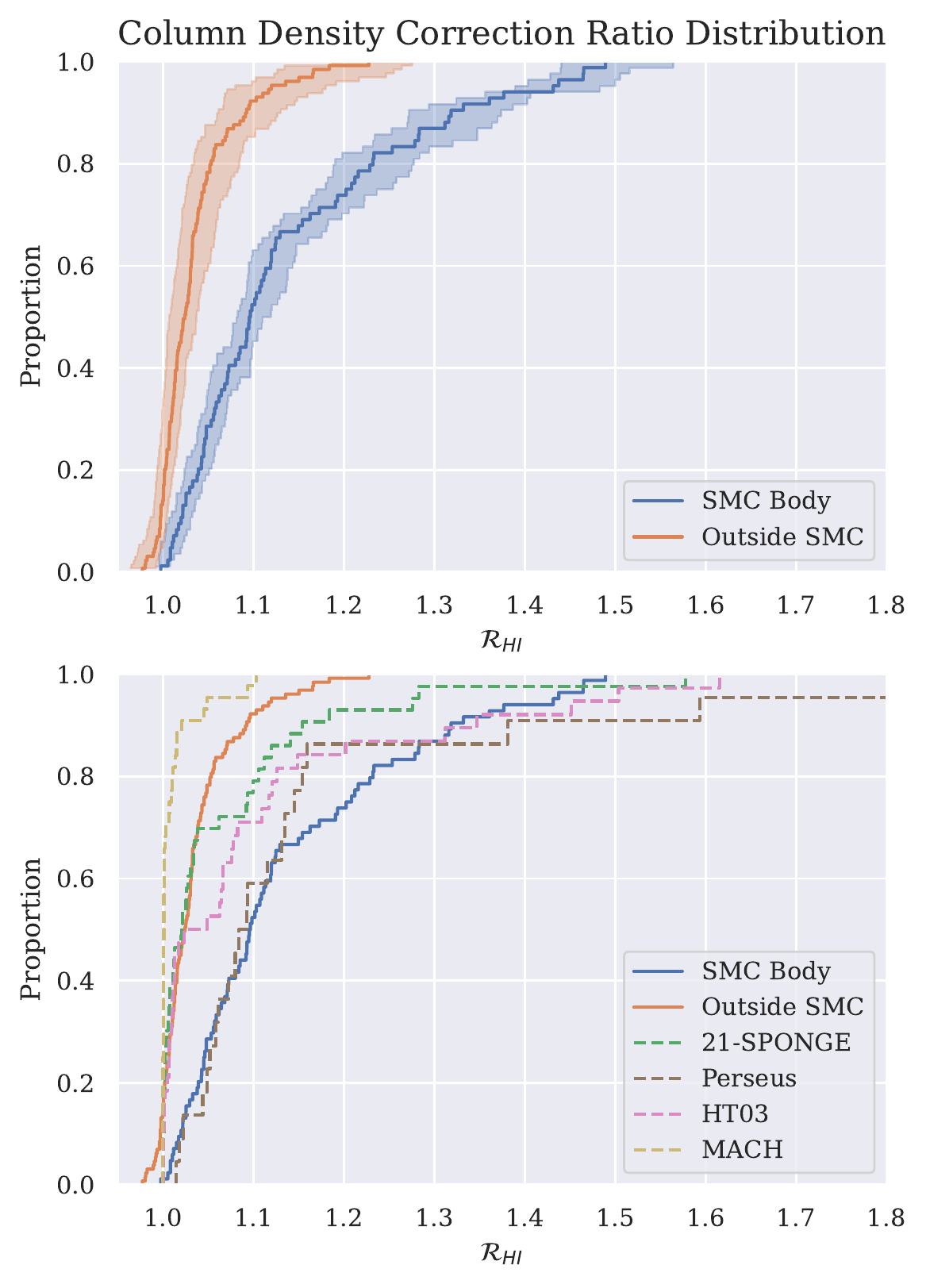}
    \caption{(top) The cumulative distribution function (CDF) with uncertainties of column density correction factor for sight-lines within the body of the SMC and outside the SMC.
    (bottom) Comparison of column density correction factor CDF for the two samples against
    \citet[][``21-SPONGE'']{Murray+2018},
    \citet[][``Perseus'']{Stanimirovic+2014},
    \citet[][``HT03'']{Heiles+Troland2003}, and \citet[][``MACH'']{Murray+2021}.}
    \label{fig:rhicdf}
\end{figure}

We compare the results for the field and the body to recent absorption studies of the Milky Way (see Fig.~\ref{fig:rhicdf}, bottom).
The field shows a similar, but slightly denser, CDF to the MACH survey \citep{Murray+2021}.
The MACH survey probed a window of very low density Galactic gas at high latitudes.
In contrast, the SMC CDF is closest to the Perseus cloud \citep{Lee+2015} at lower $\mathcal{R}_{\rm HI}$ but lacks the higher $\mathcal{R}_{\rm HI}$, likely as our sensitivity is not currently sufficient to probe the highest correction factors that they find.
Overall this suggests correction factors similar to the envelopes of Giant Molecular Clouds.

\begin{figure}[ht]
    \centering
    \includegraphics[width=\linewidth]{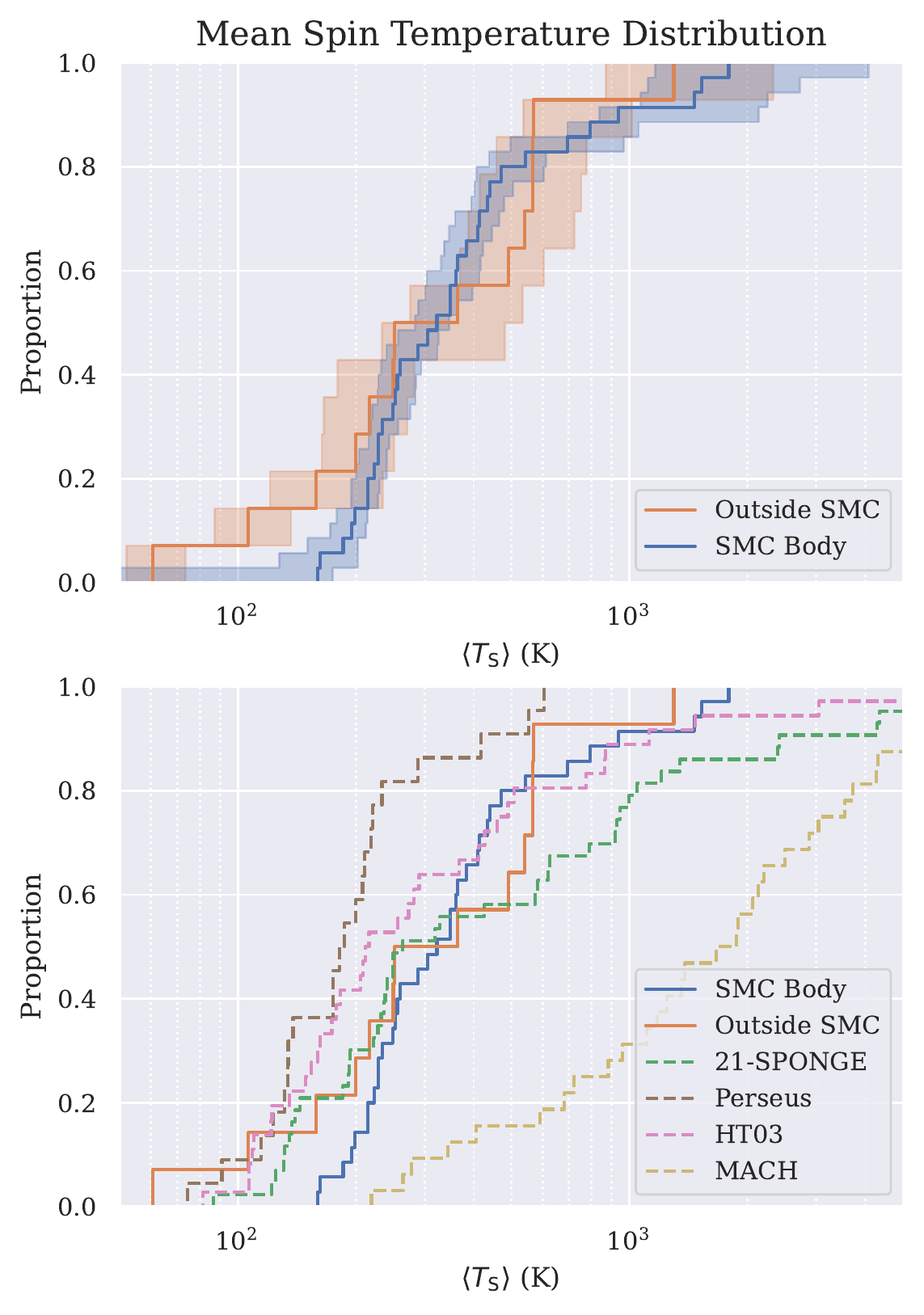}
    \caption{(top) The cumulative distribution function (CDF) with uncertainties of mean spin temperature ($\langle T_{\rm S} \rangle$) for sight-lines within the body of the SMC and outside the SMC where absorption was detected but were not saturated.
    (bottom) Comparison of mean spin temperature CDF for the two samples against
    \citet[][``21-SPONGE'']{Murray+2018},
    \citet[][``Perseus'']{Stanimirovic+2014},
    \citet[][``HT03'']{Heiles+Troland2003}, and \citet[][``MACH'']{Murray+2021}.}
    \label{fig:meantscdf}
\end{figure}

We also compare the distribution of mean spin temperatures with the same Galactic \hi\ absorption studies.
Fig.~\ref{fig:meantscdf} shows the mean spin temperatures of all sightlines with detected \hi\ absorption, but not saturated, for the SMC body and the field surrounding the SMC, along with comparisons against the recent Galactic \hi\ absorption samples.
Over 70\% of temperatures in the body of the SMC are in the $200 < \langle T_{\rm S} \rangle < 500$ K range, with a slightly lower proportion in the field.
Note that our sensitivity in this study limits our ability to detect high spin temperatures ($T_{\rm S} \ge 1150$ K).
The two distributions are still remarkably similar, likely indicating that the cold gas we detect in the field is associated with outflows from the SMC. 
Both the SMC body and the field show similar temperatures to the mid-high latitude ($|b| > 10 \degree$) HT03 \citep{Heiles+Troland2003} and 21-SPONGE \citep{Murray+2018} samples, but lacking the higher spin temperatures.
These two studies surveyed a variety of environments across the Milky Way, rather than focusing on a single region.

\subsection{Mean Spectrum of the SMC}

\begin{figure}[ht]
    \centering
    \includegraphics[width=\linewidth]{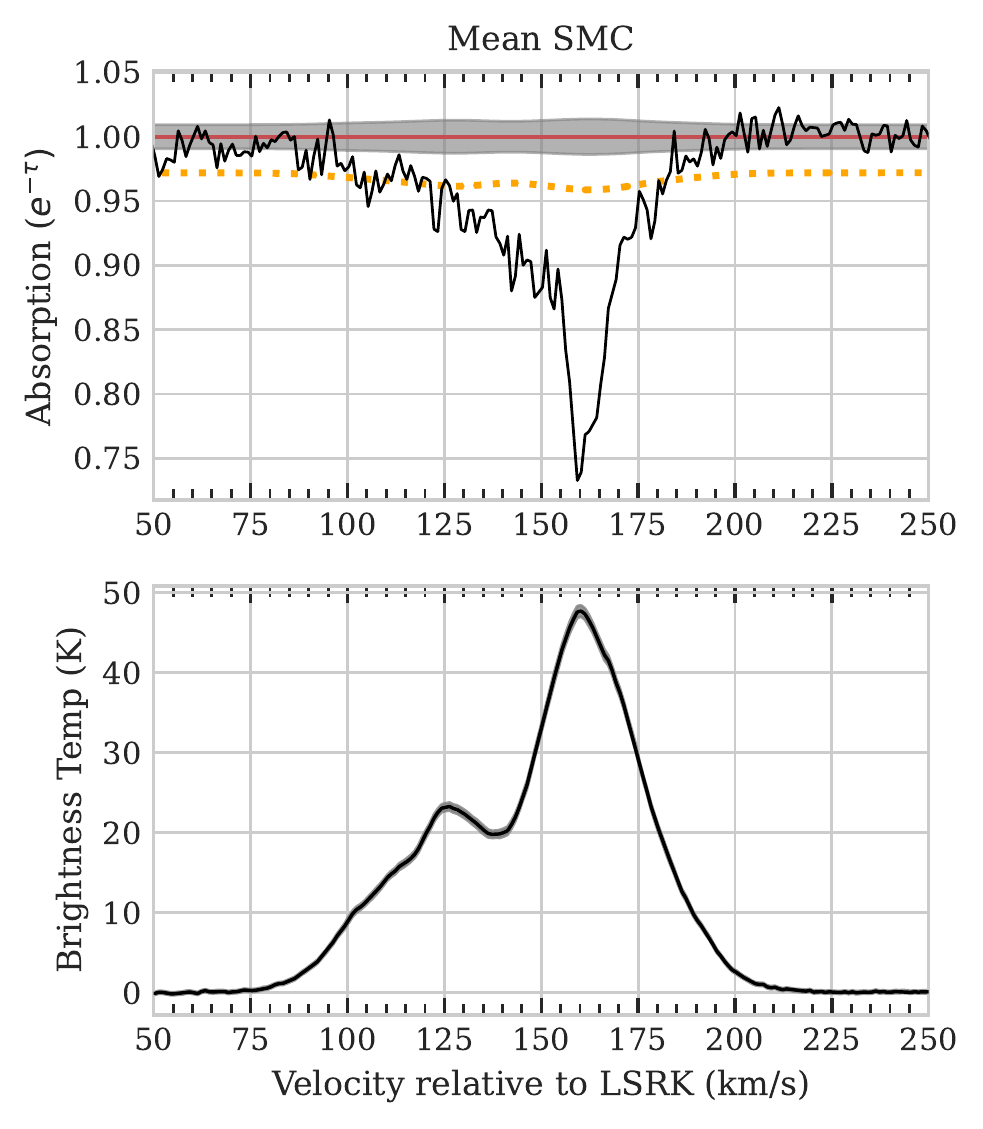}
    \caption{Weighted mean spectrum for all sight-lines through the SMC. 
    The absorption spectrum is weighted by the inverse square of the continuum noise level for each spectrum. The emission spectrum is unweighted. No correction for the SMC rotation has been applied. See Figure~\ref{fig:spectrum} for details.}
    \label{fig:meanSpectrum}
\end{figure}

The large number of sight-lines sampled through the SMC allow us to build a mean spectrum across the galaxy.
This provides a view of the where the cold gas is, and is not, across the SMC.
In Figure~\ref{fig:meanSpectrum} we show the unweighted mean of spectra for the 43 sight-lines with $\sigma_{\rm cont} \le 0.1$ within the body of the SMC.
The spectrum has not been corrected for the rotation of the SMC.
Notably, there is still fine velocity structure in the mean SMC spectrum.

At the primary emission peak ($155 \le v_{\rm LSR} \le 165$ km s$^{-1}$) 14\% of the gas is cold.
While at the secondary emission peak ($120 \le v_{\rm LSR} \le 130$ km s$^{-1}$) only 7\% of the gas is cold.
The absorption falls away much faster from the peak at 160 km s$^{-1}$ towards higher velocities, predominantly associated with the wing, than towards lower velocities, which are associated with the bar.
Overall, this mean spectrum has a total cold gas fraction of $f_{\rm c} \approx 9\%$ and a column density correction factor of $\mathcal{R}_{\rm HI} \approx 1.06$.

\subsection{Comparing velocity distributions of absorption and emission}

$21\rm\,cm$ emission in the SMC is known to exhibit a complex, multi-peaked velocity structure \citep{1999MNRAS.302..417S}. Here we investigate whether \hi\ absorption is found preferentially in one velocity component or another, or if it follows the same velocity distribution of \hi\ emission.

To identify the velocities of detected absorption components, we take the first numerical derivative of all spectra (here in the form of $1-e^{-\tau(v)}$). As the noise in the spectrum is amplified by the derivative operation, we first smooth each spectrum by a Gaussian kernel of standard deviation of two channels. Detected components are identified by their central channels ($v_{\rm c}$) if they pass the following criteria: (1) the first derivative crosses the zero line; (2) the feature is a maximum (the derivative decreases between adjacent channels); (3) the feature is detected above $3\sigma$ at $v_c$ and above $2.8\sigma$ for adjacent channels; and (4) the feature falls within the SMC velocity range ($70<v_c<250\rm\,km\,s^{-1}$). We repeat the same process for the emission spectra.

\begin{figure*}
    \centering
    \includegraphics[width=\linewidth]{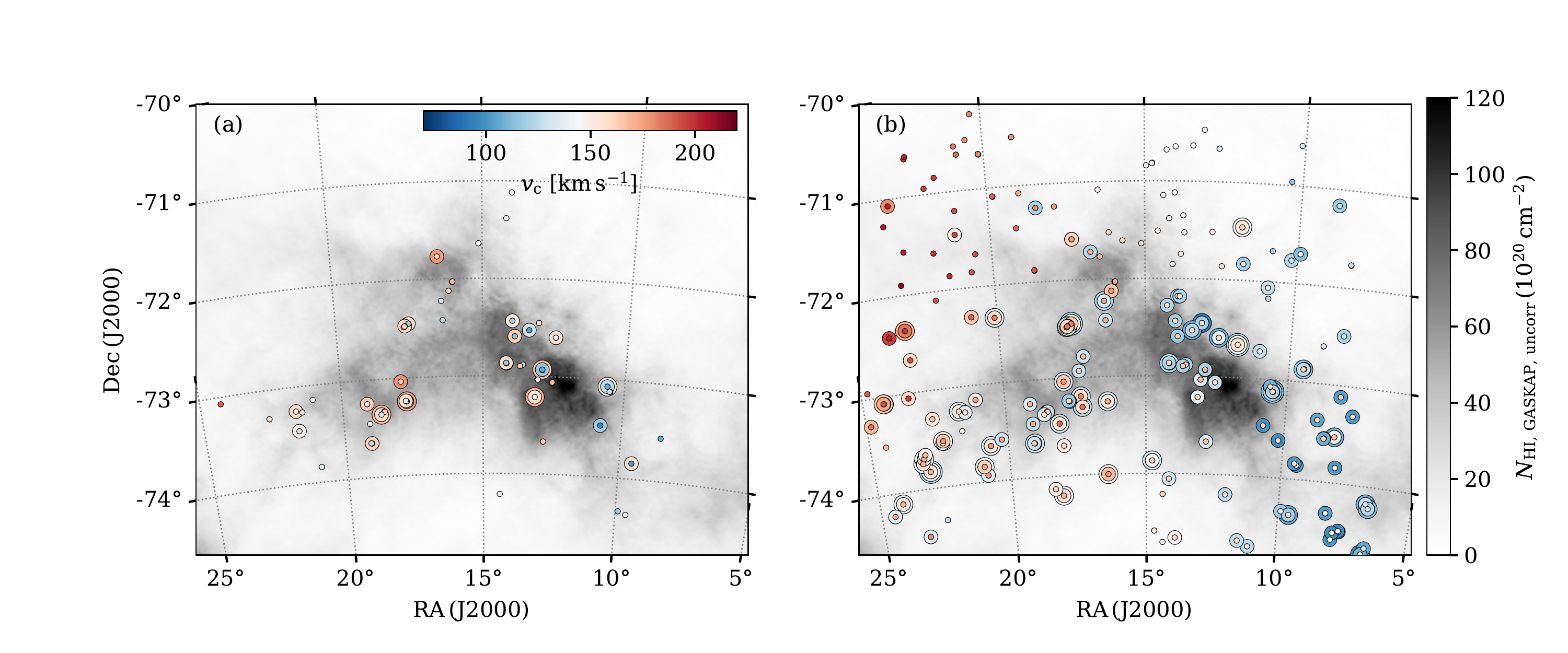}
    \caption{Comparing the positions of detected absorption (panel a) and emission (panel b) components, coloured by their central velocities ($v_c$), identified by their smoothed spectral derivatives, overlaid on an \hi\ column density map. Lines of sight featuring multiple components are shown as concentric circles. }
    \label{fig:vel_maps}
\end{figure*}

Figure~\ref{fig:vel_maps} displays maps of the positions of detected absorption (a) and emission (b) components, coloured by $v_c$. Multiple components per line of sight are displayed as concentric circles. 

\begin{figure}
    \centering
    \includegraphics[width=\linewidth]{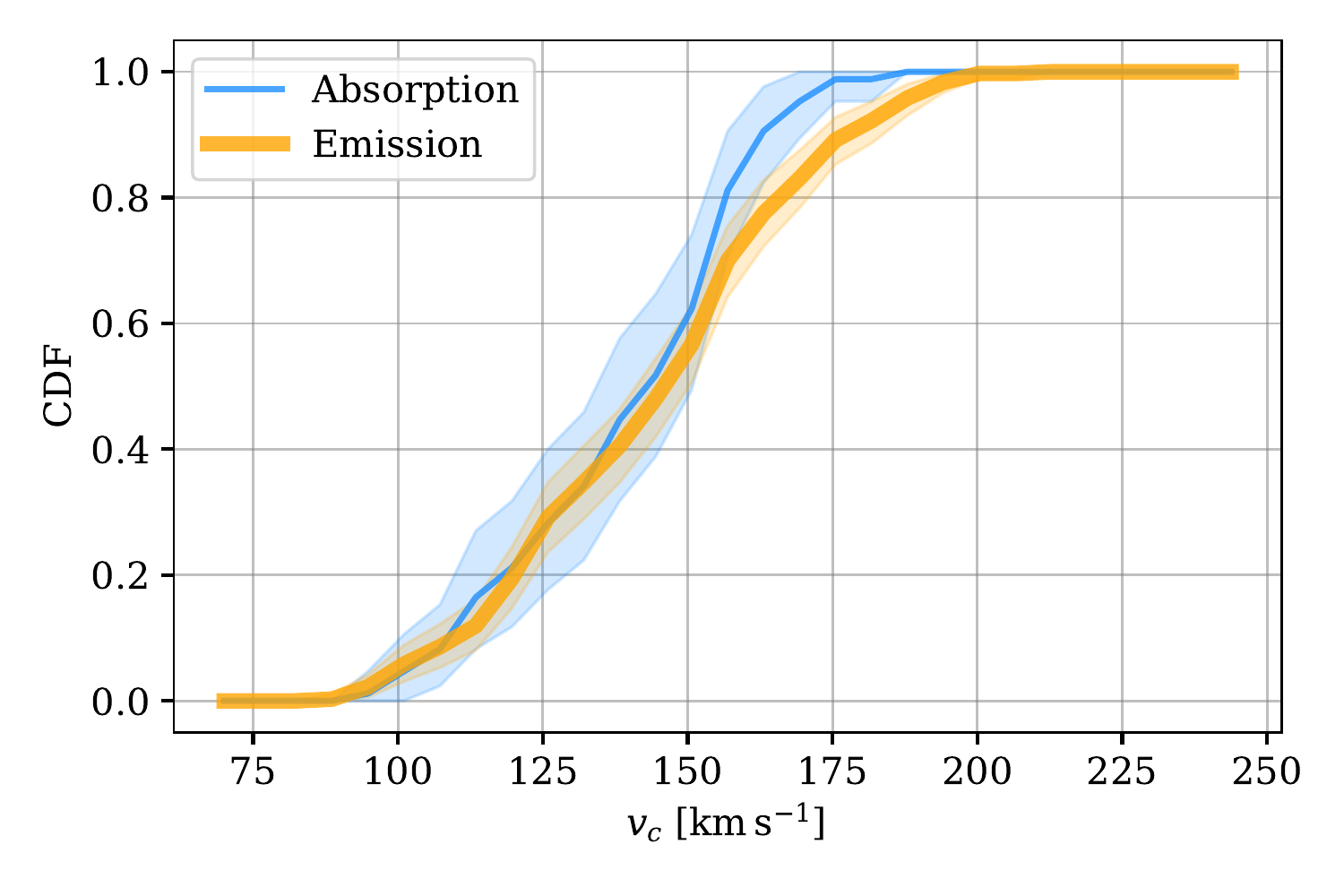}
    \caption{Cumulative distribution functions (CDFs) of the central velocities ($v_c$) of absorption (blue) and emission (orange) components. Uncertainties on the CDFs are computed by bootstrapping each sample with replacement over $10^4$ trials, and represent the $1^{\rm st}$ through $99^{\rm th}$ percentiles. }
    \label{fig:vel_cdfs}
\end{figure}

In Figure~\ref{fig:vel_cdfs}, we compare the cumulative distribution functions of $v_c$ for absorption and emission. Uncertainties on the distributions are computed by bootstrapping each sample with replacement over $10^4$ trials, and represent the $1^{\rm st}$ through $99^{\rm th}$ percentiles. Although the distributions are indistinguishable between $100<v_c<150\rm km\,s^{-1}$, absorption components are not found in these data beyond $v_c \sim 175\rm\,km\,s^{-1}$ whereas emission components extend to $v_c>200\rm\,km\,s^{-1}$.
Using a two-sided Kolmogorov–Smirnov test across $10^4$ bootstrapped trials we find a p-value$\sim0.17$.
This preliminary result may indicate that the absorption components are more likely to be found in low-velocity gas in the SMC relative to emission components.
Future, deeper GASKAP observations of the SMC will provide the opportunity to examine this result further.

\section{Conclusion}

In this paper we have presented the first untargeted survey of \hi\ absorption in the SMC, alongside the pipeline used to process these and future GASKAP \hi\ absorption data.
We have produced \hi\ absorption spectra against 337 continuum sources, matched those to emission spectra from the \cite{pingel2021} SMC emission cube, and analysed 229 of those which meet our quality requirements.
This represents a 275\% increase on sight-lines over previous studies of the SMC region.
We have also described the major choices for the pipeline and demonstrated the effectiveness of the pipeline using the SMC data as an example.

We have found 122 absorption features across 65 of the spectra, with no features detected in the other 164 spectra.
Within the body of the SMC ($N_{\rm HI,uncorr} > 2 \times 10^{21}$ cm$^{-1}$) we found absorption features against 49 of 79 (62\%) spectra, including in all sight-lines with $\rm{S_{cont}} \ge 100$ mJy.
Outside the SMC body we found absorption features against only 16 of 150 (11\%) of spectra.

From this first work exploring GASKAP-HI's unbiased view of absorption across the SMC and its surrounds, we have made two primary findings:

\begin{itemize}
    \item The median fraction of \hi\ in the SMC in the CNM is $f_{\rm c} \approx 11\%$, which is lower than found in \cite{Jameson+2019} but higher than the fraction reported in \cite{Dickey+00}.
This value is driven by the observation of more shallow absorption in less dense regions.
This new fraction better represents the cold gas across the whole of the SMC, rather than just the denser regions observed in \cite{Jameson+2019}.
However, it is potentially still an over estimate due to our sensitivity limits on detection of shallow absorption.
Future more sensitive GASKAP-HI observations of the field will refine this figure.

    \item The range of column density correction factors for optically thick gas varies greatly between the wing and the bar of the SMC. 
The bar shows a linear increase in correction factor with column density ($\mathcal{R}_{\rm HI} \simeq 1 + 0.51[\text{log}_{10}N - 21.43]$) whilst the wing shows a large variety of correction values with no relationship to uncorrected column density.
Overall, the SMC is closer in optical thickness to Galactic molecular clouds, such as the Perseus cloud, whilst the field surrounding the SMC is similar to low density fields at high Galactic latitudes.
\end{itemize}

In addition we have found that:
\begin{enumerate}[label=\alph*)]
    \item The SMC has an inverse noise weighted mean spin temperature $\langle T_{\rm S} \rangle = 245$ K; which is lower than that found by \cite{Dickey+00}
    \item There is no significant trend of mean spin temperature with column density
    \item The equivalent widths and uncorrected column densities we see in the SMC are predominantly within the Milky Way cold gas formation thresholds proposed by \cite{Kanekar+2011}, however we see a higher limit for the condensation of dense \hi\ into molecular H$_2$ saturation in the SMC, due to 
    the lower metallicity of the SMC, and we see hints of cold gas formation below their threshold
    \item Fine velocity structure is present even in the mean SMC spectrum, which implies that this structure may also be seen in extra-Galactic experiments
    \item There are indications that absorption components may be preferentially found in the lower velocity gas of the SMC as compared to the emission components.
\end{enumerate}

Further exploration of the pilot SMC absorption data is planned in future GASKAP-HI papers.
In addition, a much longer observation (up to 200 hrs) of the SMC in the full GASKAP-HI survey will obtain \hi\ absorption spectra up to 3 times as sensitive as the spectra presented here.
We will have the opportunity to explore the column density threshold for the formation of cold \hi\ proposed by \cite{Kanekar+2011} in more detail.
We also expect that we will see further clear detections of shallow absorption features, and thus slightly refine the cold gas fraction for the SMC.

\begin{acknowledgement}
The Parkes radio telescope is part of the Australia Telescope National Facility which is funded by the Australian Government for operation as a National Facility managed by CSIRO.
The Australian SKA Pathfinder is part of the Australia Telescope National Facility which is managed by CSIRO. 
Operation of ASKAP is funded by the Australian Government with support from the National Collaborative Research Infrastructure Strategy. 
ASKAP uses the resources of the Pawsey Supercomputing Centre. 
Establishment of ASKAP, the Murchison Radio-astronomy Observatory and the Pawsey Supercomputing Centre are initiatives of the Australian Government, with support from the Government of Western Australia and the Science and Industry Endowment Fund. 
We acknowledge the Wajarri Yamatji people as the traditional owners of the Observatory site.
This paper includes archived data obtained through the CSIRO ASKAP Science Data Archive, CASDA (http://data.csiro.au).

This research was supported by the Australian Research Council (ARC) through grant DP190101571.
N.M.-G. acknowledges the support of the ARC through Future Fellowship FT150100024.
S.S. gratefully acknowledges support by the NSF grant~AST-2108370.
We thank the anonymous reviewer for their thoughtful comments that have improved this paper.
\end{acknowledgement}

\bibliography{main}{}

\end{document}

%% file: spectra_table.tex
\begin{tabular}{rcrrrcrrrrr}
\hline \hline
ID & Source & RA & Dec & Peak Flux & Rating & $\sigma_{\rm cont}$ & Peak $\tau$ & N$_{\rm HI,uncorr}$ & $\langle T_{\rm S} \rangle$ & $\mathcal{R}_{\rm HI}$ \\
 &  & (deg) & (deg) & (mJy) &  &  &  & ($10^{21}$ cm$^{-2}$) &  &  \\
\hline
178 & J004718$-$723947 & 11.8285 & $-72.663$ & 32 & B & 0.133 & $0.99\pm0.43$ & 4.0 & $265^{+76}_{-45}$ & $1.10^{+0.02}_{-0.02}$ \\
304 & J004956$-$723554 & 12.4863 & $-72.599$ & 82 & A & 0.034 & $0.72\pm0.10$ & 4.2 & $321^{+26}_{-21}$ & $1.08^{+0.01}_{-0.00}$ \\
33 & J005014$-$730326 & 12.5617 & $-73.057$ & 17 & C & 0.197 & $>1.12$ & 10.3 & $482^{+147}_{-97}$ & $1.23^{+0.04}_{-0.05}$ \\
74 & J005141$-$725545 & 12.9231 & $-72.929$ & 17 & B & 0.136 & $>1.45$ & 8.6 & $260^{+32}_{-26}$ & $1.31^{+0.05}_{-0.04}$ \\
75 & J005141$-$725603 & 12.9216 & $-72.934$ & 19 & B & 0.137 & $>1.44$ & 8.6 & $251^{+35}_{-24}$ & $1.32^{+0.05}_{-0.04}$ \\
62 & J005217$-$730157 & 13.0714 & $-73.033$ & 18 & B & 0.146 & $1.28\pm0.63$ & 7.8 & $291^{+47}_{-39}$ & $1.22^{+0.04}_{-0.04}$ \\
357 & J005218$-$722708 & 13.0779 & $-72.452$ & 153 & B & 0.019 & $0.22\pm0.04$ & 3.9 & $937^{+131}_{-94}$ & $1.02^{+0.00}_{-0.00}$ \\
319 & J005238$-$731245 & 13.1589 & $-73.213$ & 101 & A & 0.035 & $3.26\pm0.92$ & 8.9 & $234^{+7}_{-6}$ & $1.38^{+0.03}_{-0.02}$ \\
287 & J005337$-$723144 & 13.4067 & $-72.529$ & 69 & A & 0.047 & $>2.61$ & 5.0 & $232^{+15}_{-13}$ & $1.12^{+0.02}_{-0.01}$ \\
200 & J005423$-$725257 & 13.5979 & $-72.883$ & 35 & B & 0.105 & $>1.68$ & 7.6 & $405^{+79}_{-53}$ & $1.15^{+0.03}_{-0.02}$ \\
19 & J005448$-$725353 & 13.7026 & $-72.898$ & 16 & C & 0.181 & $>1.18$ & 7.4 & $224^{+40}_{-28}$ & $1.32^{+0.06}_{-0.05}$ \\
13 & J005535$-$723533 & 13.8994 & $-72.593$ & 15 & C & 0.162 & $>1.38$ & 7.2 & $529^{+206}_{-125}$ & $1.23^{+0.05}_{-0.04}$ \\
221 & J005642$-$725216 & 14.1764 & $-72.871$ & 25 & C & 0.125 & $>1.58$ & 7.8 & $282^{+41}_{-32}$ & $1.25^{+0.04}_{-0.03}$ \\
222 & J005644$-$725200 & 14.1872 & $-72.867$ & 39 & C & 0.087 & $>1.94$ & 8.0 & $279^{+24}_{-23}$ & $1.21^{+0.03}_{-0.02}$ \\
\hline
\end{tabular}

%% file: absorption_table.tex
\begin{tabular}{llrrrrrrr}
\hline \hline
Source & Feature & Min Velocity & Max Velocity & Width & Peak & Peak $\tau$ & Significance & Equiv. Width \\
 &  & (kms$^{-1}$) & (kms$^{-1}$) & (kms$^{-1}$) & Absorption &  & (sigma) & (kms$^{-1}$) \\
\hline
J004956$-$723554 & J004956$-$723554\_$-$2 & $-2.3$ & $-0.3$ & 3 & $0.16\pm0.04$ & $0.18\pm0.05$ & 4.1 & $0.45\pm0.07$ \\
J004956$-$723554 & J004956$-$723554\_137 & $137.7$ & $150.7$ & 14 & $0.51\pm0.05$ & $0.72\pm0.10$ & 10.1 & $5.37\pm0.20$ \\
J004956$-$723554 & J004956$-$723554\_155 & $155.7$ & $160.7$ & 6 & $0.45\pm0.05$ & $0.60\pm0.09$ & 8.7 & $2.13\pm0.13$ \\
J004956$-$723554 & J004956$-$723554\_3 & $3.7$ & $4.7$ & 2 & $0.18\pm0.04$ & $0.19\pm0.05$ & 4.6 & $0.38\pm0.06$ \\
J005014$-$730326 & J005014$-$730326\_152 & $152.9$ & $153.9$ & 2 & $1.06\pm0.32$ & $>1.14$ & 3.3 & $2.28\pm0.54$ \\
J005141$-$725545 & J005141$-$725545\_106 & $106.8$ & $108.8$ & 3 & $0.71\pm0.19$ & $1.24\pm0.51$ & 3.8 & $3.18\pm0.37$ \\
J005141$-$725545 & J005141$-$725545\_114 & $114.8$ & $117.8$ & 4 & $1.03\pm0.24$ & $>1.45$ & 4.5 & $6.55\pm0.52$ \\
J005141$-$725603 & J005141$-$725603\_113 & $113.8$ & $120.8$ & 8 & $1.19\pm0.24$ & $>1.44$ & 5.0 & $15.96\pm0.76$ \\
J005218$-$722708 & J005218$-$722708\_0 & $-0.4$ & $1.6$ & 3 & $0.13\pm0.02$ & $0.14\pm0.03$ & 5.8 & $0.31\pm0.04$ \\
J005218$-$722708 & J005218$-$722708\_153 & $153.6$ & $159.6$ & 7 & $0.20\pm0.03$ & $0.22\pm0.04$ & 6.5 & $0.98\pm0.08$ \\
J005218$-$722708 & J005218$-$722708\_161 & $161.6$ & $162.6$ & 2 & $0.12\pm0.03$ & $0.13\pm0.03$ & 4.0 & $0.21\pm0.04$ \\
J005238$-$731245 & J005238$-$731245\_0 & $-0.0$ & $1.0$ & 2 & $0.20\pm0.04$ & $0.22\pm0.05$ & 4.6 & $0.35\pm0.06$ \\
J005238$-$731245 & J005238$-$731245\_132 & $133.0$ & $166.0$ & 34 & $0.96\pm0.06$ & $3.26\pm0.92$ & 16.5 & $35.60\pm0.35$ \\
J005238$-$731245 & J005238$-$731245\_169 & $170.0$ & $174.0$ & 5 & $0.31\pm0.05$ & $0.37\pm0.07$ & 6.2 & $1.47\pm0.12$ \\
J005337$-$723144 & J005337$-$723144\_102 & $102.6$ & $107.6$ & 6 & $0.27\pm0.06$ & $0.31\pm0.07$ & 4.7 & $1.40\pm0.15$ \\
J005337$-$723144 & J005337$-$723144\_122 & $122.6$ & $124.6$ & 3 & $0.35\pm0.08$ & $0.44\pm0.12$ & 4.4 & $1.21\pm0.14$ \\
J005337$-$723144 & J005337$-$723144\_127 & $127.6$ & $134.6$ & 8 & $1.05\pm0.07$ & $>2.61$ & 14.3 & $9.35\pm0.21$ \\
J005423$-$725257 & J005423$-$725257\_122 & $122.8$ & $125.8$ & 4 & $1.18\pm0.19$ & $>1.68$ & 6.3 & $4.87\pm0.41$ \\
J005423$-$725257 & J005423$-$725257\_174 & $174.8$ & $176.8$ & 3 & $0.69\pm0.15$ & $1.17\pm0.38$ & 4.7 & $2.74\pm0.27$ \\
J005448$-$725353 & J005448$-$725353\_116 & $116.8$ & $118.8$ & 3 & $1.12\pm0.31$ & $>1.17$ & 3.6 & $7.67\pm0.63$ \\
J005448$-$725353 & J005448$-$725353\_163 & $163.8$ & $167.8$ & 5 & $1.48\pm0.31$ & $>1.18$ & 4.8 & $11.32\pm0.82$ \\
J005535$-$723533 & J005535$-$723533\_113 & $113.6$ & $114.6$ & 2 & $1.33\pm0.25$ & $>1.38$ & 5.3 & $5.98\pm0.42$ \\
J005535$-$723533 & J005535$-$723533\_116 & $116.6$ & $117.6$ & 2 & $1.07\pm0.27$ & $>1.31$ & 4.0 & $2.83\pm0.45$ \\
J005642$-$725216 & J005642$-$725216\_158 & $158.7$ & $160.7$ & 3 & $1.11\pm0.21$ & $>1.58$ & 5.4 & $5.53\pm0.40$ \\
J005644$-$725200 & J005644$-$725200\_117 & $117.7$ & $118.7$ & 2 & $0.50\pm0.15$ & $0.69\pm0.25$ & 3.4 & $1.30\pm0.22$ \\
J005644$-$725200 & J005644$-$725200\_155 & $155.7$ & $160.7$ & 6 & $1.09\pm0.14$ & $>1.94$ & 7.6 & $6.66\pm0.38$ \\
J005644$-$725200 & J005644$-$725200\_171 & $171.7$ & $172.7$ & 2 & $0.51\pm0.14$ & $0.71\pm0.25$ & 3.7 & $1.32\pm0.21$ \\
J005644$-$725200 & J005644$-$725200\_175 & $175.7$ & $177.7$ & 3 & $0.57\pm0.13$ & $0.84\pm0.26$ & 4.5 & $1.86\pm0.23$ \\
\hline
\end{tabular}

%% file: main.bbl
\begin{thebibliography}{}
\expandafter\ifx\csname natexlab\endcsname\relax\def\natexlab#1{#1}\fi

\bibitem[{Bialy \& Sternberg(2016)}]{Bialy+16}
Bialy, S., \& Sternberg, A. 2016, ApJ, 822, 83

\bibitem[{Bialy \& Sternberg(2019)}]{Bialy+2019}
---. 2019, ApJ, 881, 160

\bibitem[{Bolatto {et~al.}(2011)Bolatto, Leroy, Jameson, Ostriker, Gordon,
  Lawton, Stanimirović, Israel, Madden, Hony, Sandstrom, Bot, Rubio, Winkler,
  Roman-Duval, Loon, Oliveira, \& Indebetouw}]{Bolatto+11}
Bolatto, A.~D., Leroy, A.~K., Jameson, K., {et~al.} 2011, ApJ, 741, 12

\bibitem[{Brown {et~al.}(2014)Brown, Dickey, Dawson, \&
  McClure-Griffiths}]{2014ApJS..211...29B}
Brown, C., Dickey, J.~M., Dawson, J.~R., \& McClure-Griffiths, N.~M. 2014,
  ApJS, 211, 29

\bibitem[{Chengalur {et~al.}(2013)Chengalur, Kanekar, \& Roy}]{Chengalur+2013}
Chengalur, J.~N., Kanekar, N., \& Roy, N. 2013, \mnras, 432, 3074

\bibitem[{Cornwell \& Perley(1992)}]{1992a&a...261..353c}
Cornwell, T., \& Perley, R. 1992, A\&A, 261, 353

\bibitem[{Dempsey(2022)}]{james_dempsey_2022_6388108}
Dempsey, J. 2022, GASKAP/GASKAP-HI-Absorption-Pipeline: v1.0.0,
  doi:\url{10.5281/zenodo.6388108}

\bibitem[{Dempsey {et~al.}(2022)Dempsey, {Mcclure-Griffiths}, {Murray},
  {Dickey}, {Pingel}, \& {Jameson}}]{Dempsey2021}
Dempsey, J., {Mcclure-Griffiths}, N., {Murray}, C., {et~al.} 2022, GASKAP-HI
  Pilot Survey Science - Small Magellanic Cloud HI absorption spectra,
  \url{https://doi.org/10.25919/1va2-h045}, doi:\url{10.25919/1va2-h045}

\bibitem[{Dempsey {et~al.}(2020)Dempsey, McClure-Griffiths, Jameson, \&
  Buckland-Willis}]{2020MNRAS.496..913D}
Dempsey, J., McClure-Griffiths, N.~M., Jameson, K., \& Buckland-Willis, F.
  2020, \mnras, 496, 913

\bibitem[{Dickey \& Benson(1982)}]{DB82}
Dickey, J.~M., \& Benson, J.~M. 1982, AJ, 87, 278

\bibitem[{Dickey {et~al.}(1992)Dickey, Brinks, \& Puche}]{1992ApJ...385..501D}
Dickey, J.~M., Brinks, E., \& Puche, D. 1992, ApJ, 385, 501

\bibitem[{Dickey \& Lockman(1990)}]{Dickey+Lockman90}
Dickey, J.~M., \& Lockman, F.~J. 1990, ARA\&A, 28, 215

\bibitem[{Dickey {et~al.}(2000)Dickey, Mebold, Stanimirović, \&
  Staveley-Smith}]{Dickey+00}
Dickey, J.~M., Mebold, U., Stanimirović, S., \& Staveley-Smith, L. 2000, ApJ,
  536, 756

\bibitem[{Dickey {et~al.}(2013)Dickey, McClure-Griffiths, Gibson, Gomez, Imai,
  Jones, Stanimirovic, van Loon, Walsh, Alberdi, Anglada, Uscanga, Arce,
  Bailey, Begum, Wakker, Bekhti, Kalberla, Winkel, Bekki, For, Staveley-Smith,
  Westmeier, Burton, Cunningham, Dawson, Ellingsen, Diamond, Green, Hill,
  Koribalski, McConnell, Rathborne, Voronkov, Douglas, English, Ford, Foster,
  Gomez, Green, Bland-Hawthorn, Gulyaev, Hoare, Joncas, Kang, Kerton, Koo,
  Leahy, Lo, Lockman, Migenes, Nakashima, Zhang, Nidever, Peek, Tafoya, Tian,
  \& Wu}]{2013PASA...30....3D}
Dickey, J.~M., McClure-Griffiths, N., Gibson, S.~J., {et~al.} 2013, PASA, 30,
  e003

\bibitem[{Field(1958)}]{Field58}
Field, G.~B. 1958, Proceedings of the IRE, 46, 240

\bibitem[{Graczyk {et~al.}(2014)Graczyk, Pietrzyński, Thompson, Gieren,
  Pilecki, Konorski, Udalski, Soszyński, Villanova, Górski, Suchomska,
  Karczmarek, Kudritzki, Bresolin, \& Gallenne}]{Graczyk+2014}
Graczyk, D., Pietrzyński, G., Thompson, I.~B., {et~al.} 2014, ApJ, 780, 59

\bibitem[{Heiles \& Troland(2003)}]{Heiles+Troland2003}
Heiles, C., \& Troland, T.~H. 2003, ApJS, 145, 329

\bibitem[{Hennebelle \& Audit(2007)}]{Hennebelle+2007}
Hennebelle, P., \& Audit, E. 2007, A\&A, 465, 431

\bibitem[{Hotan {et~al.}(2021)Hotan, Bunton, Chippendale, Whiting, Tuthill,
  Moss, McConnell, Amy, Huynh, Allison, Anderson, Bannister, Bastholm,
  Beresford, Bock, Bolton, Chapman, Chow, Collier, Cooray, Cornwell, Diamond,
  Edwards, Feain, Franzen, George, Gupta, Hampson, Harvey-Smith, Hayman,
  Heywood, Jacka, Jackson, Jackson, Jeganathan, Johnston, Kesteven, Kleiner,
  Koribalski, Lee-Waddell, Lenc, Lensson, Mackay, Mahony, McClure-Griffiths,
  McConigley, Mirtschin, Ng, Norris, Pearce, Phillips, Pilawa, Raja, Reynolds,
  Roberts, Roxby, Sadler, Shields, Schinckel, Serra, Shaw, Sweetnam, Troup,
  Tzioumis, Voronkov, \& Westmeier}]{2021PASA...38....9H}
Hotan, A.~W., Bunton, J.~D., Chippendale, A.~P., {et~al.} 2021, PASA, 38, e009

\bibitem[{Jameson {et~al.}(2019)Jameson, McClure-Griffiths, Liu, Dickey,
  Staveley-Smith, Stanimirović, Dempsey, Dawson, Dénes, Bolatto, Li, \&
  Wong}]{Jameson+2019}
Jameson, K.~E., McClure-Griffiths, N.~M., Liu, B., {et~al.} 2019, ApJS, 244, 7

\bibitem[{Kalberla \& Haud(2015)}]{2015A&A...578A..78K}
Kalberla, P. M.~W., \& Haud, U. 2015, A\&A, 578, A78

\bibitem[{Kanekar {et~al.}(2011)Kanekar, Braun, \& Roy}]{Kanekar+2011}
Kanekar, N., Braun, R., \& Roy, N. 2011, ApJL, 737, L33

\bibitem[{{Kennicutt} \& {Evans}(2012)}]{Kennicutt+2012}
{Kennicutt}, R.~C., \& {Evans}, N.~J. 2012, ARA\&A, 50, 531

\bibitem[{Kim \& Ostriker(2017)}]{Kim+2017}
Kim, C.-G., \& Ostriker, E.~C. 2017, ApJ, 846, 133

\bibitem[{Krumholz {et~al.}(2009)Krumholz, McKee, \& Tumlinson}]{Krumholz+2009}
Krumholz, M.~R., McKee, C.~F., \& Tumlinson, J. 2009, ApJ, 693, 216

\bibitem[{Lee {et~al.}(2015)Lee, Stanimirović, Murray, Heiles, \&
  Miller}]{Lee+2015}
Lee, M.-Y., Stanimirović, S., Murray, C.~E., Heiles, C., \& Miller, J. 2015,
  ApJ, 809, 56

\bibitem[{Liszt(2001)}]{Liszt01}
Liszt, H. 2001, A\&A, 371, 698

\bibitem[{McClure-Griffiths {et~al.}(2009)McClure-Griffiths, Pisano,
  Calabretta, Ford, Lockman, Staveley-Smith, Kalberla, Bailin, Dedes,
  Janowiecki, Gibson, Murphy, Nakanishi, \& Newton-McGee}]{2009ApJS..181..398M}
McClure-Griffiths, N.~M., Pisano, D.~J., Calabretta, M.~R., {et~al.} 2009,
  ApJS, 181, 398

\bibitem[{McClure-Griffiths {et~al.}(2018)McClure-Griffiths, Dénes, Dickey,
  Stanimirovic, Staveley-Smith, Jameson, Teodoro, Allison, Collier,
  Chippendale, Franzen, Gürkan, Heald, Hotan, Kleiner, Lee-Waddell, McConnell,
  Popping, Rhee, Riseley, Voronkov, \& Whiting}]{McClure-Griffiths+18}
McClure-Griffiths, N.~M., Dénes, H., Dickey, J.~M., {et~al.} 2018, Nature
  Astronomy, 2, 901

\bibitem[{McKee \& Ostriker(1977)}]{McKee+77}
McKee, C.~F., \& Ostriker, J.~P. 1977, ApJ, 218, 148

\bibitem[{McMullin {et~al.}(2007)McMullin, Waters, Schiebel, Young, \&
  Golap}]{2007aspc..376..127m}
McMullin, J., Waters, B., Schiebel, D., Young, W., \& Golap, K. 2007, in
  Astronomical Society of the Pacific Conference Series, Vol. 376, Astronomical
  Data Analysis Software and Systems XVI, 127

\bibitem[{Murray {et~al.}(2018)Murray, Stanimirović, Goss, Heiles, Dickey,
  Babler, \& Kim}]{Murray+2018}
Murray, C.~E., Stanimirović, S., Goss, W.~M., {et~al.} 2018, ApJS, 238, 14

\bibitem[{Murray {et~al.}(2021)Murray, Stanimirović, Heiles, Dickey,
  McClure-Griffiths, Lee, Goss, \& Killerby-Smith}]{Murray+2021}
Murray, C.~E., Stanimirović, S., Heiles, C., {et~al.} 2021, ApJS, 256, 37

\bibitem[{Murray {et~al.}(2015)Murray, Stanimirović, Goss, Dickey, Heiles,
  Lindner, Babler, Pingel, Lawrence, Jencson, \& Hennebelle}]{Murray+2015}
Murray, C.~E., Stanimirović, S., Goss, W.~M., {et~al.} 2015, ApJ, 804, 89

\bibitem[{Nguyen {et~al.}(2019)Nguyen, Dawson, Lee, Murray, Stanimirović,
  Heiles, Miville-Deschênes, \& Petzler}]{Nguyen+2019}
Nguyen, H., Dawson, J.~R., Lee, M.-Y., {et~al.} 2019, ApJ, 880, 141

\bibitem[{{Pingel} {et~al.}(2022){Pingel}, {Dempsey}, {McClure-Griffiths},
  {Dickey}, {Jameson}, {Arce}, {Anglada}, {Bland-Hawthorn}, {Breen},
  {Buckland-Willis}, {Clark}, {Dawson}, {D{\'e}nes}, {Di Teodoro}, {For},
  {Foster}, {G{\'o}mez}, {Imai}, {Joncas}, {Kim}, {Lee}, {Lynn}, {Leahy}, {Ma},
  {Marchal}, {McConnell}, {Miville-Desch{\`e}nes}, {Moss}, {Murray}, {Nidever},
  {Peek}, {Stanimirovi{\'c}}, {Staveley-Smith}, {Tepper-Garcia}, {Tremblay},
  {Uscanga}, {van Loon}, {V{\'a}zquez-Semadeni}, {Allison}, {Anderson}, {Ball},
  {Bell}, {Bock}, {Bunton}, {Cooray}, {Cornwell}, {Koribalski}, {Gupta},
  {Hayman}, {Harvey-Smith}, {Lee-Waddell}, {Ng}, {Phillips}, {Voronkov},
  {Westmeier}, \& {Whiting}}]{pingel2021}
{Pingel}, N.~M., {Dempsey}, J., {McClure-Griffiths}, N.~M., {et~al.} 2022,
  \pasa, 39, e005

\bibitem[{Russell \& Dopita(1992)}]{Russell+1992}
Russell, S.~C., \& Dopita, M.~A. 1992, ApJ, 384, 508

\bibitem[{{Stanimirovi{\'c}} {et~al.}(2007){Stanimirovi{\'c}}, {Heiles}, \&
  {Kanekar}}]{Stanimirovic+2007}
{Stanimirovi{\'c}}, S., {Heiles}, C., \& {Kanekar}, N. 2007, in Astronomical
  Society of the Pacific Conference Series, Vol. 365, SINS - Small Ionized and
  Neutral Structures in the Diffuse Interstellar Medium, ed. M.~{Haverkorn} \&
  W.~M. {Goss}, 22

\bibitem[{Stanimirovic {et~al.}(1999)Stanimirovic, Staveley-Smith, Dickey,
  Sault, \& Snowden}]{Stanimirovic+99}
Stanimirovic, S., Staveley-Smith, L., Dickey, J.~M., Sault, R.~J., \& Snowden,
  S.~L. 1999, \mnras, 302, 417

\bibitem[{{Stanimirovic} {et~al.}(1999){Stanimirovic}, {Staveley-Smith},
  {Dickey}, {Sault}, \& {Snowden}}]{1999MNRAS.302..417S}
{Stanimirovic}, S., {Staveley-Smith}, L., {Dickey}, J.~M., {Sault}, R.~J., \&
  {Snowden}, S.~L. 1999, \mnras, 302, 417

\bibitem[{Stanimirović {et~al.}(2014)Stanimirović, Murray, Lee, Heiles, \&
  Miller}]{Stanimirovic+2014}
Stanimirović, S., Murray, C.~E., Lee, M.-Y., Heiles, C., \& Miller, J. 2014,
  ApJ, 793, 132

\bibitem[{Whiting \& Humphreys(2012)}]{2012PASA...29..371W}
Whiting, M., \& Humphreys, B. 2012, PASA, 29, 371

\bibitem[{Wolfire {et~al.}(1995)Wolfire, Hollenbach, McKee, Tielens, \&
  Bakes}]{1995ApJ...443..152W}
Wolfire, M.~G., Hollenbach, D., McKee, C.~F., Tielens, A. G. G.~M., \& Bakes,
  E. L.~O. 1995, ApJ, 443, 152

\bibitem[{Wolfire {et~al.}(2003)Wolfire, McKee, Hollenbach, \&
  Tielens}]{2003ApJ...587..278W}
Wolfire, M.~G., McKee, C.~F., Hollenbach, D., \& Tielens, A. G. G.~M. 2003,
  ApJ, 587, 278

\end{thebibliography}
